\documentclass{JHEP} 

\usepackage{epsfig}

\newcommand\fverb{\setbox\pippobox=\hbox\bgroup\verb}
\newcommand\fverbdo{\egroup\medskip\noindent%
			\fbox{\unhbox\pippobox}\ }
\newcommand\fverbit{\egroup\item[\fbox{\unhbox\pippobox}]}
\newbox\pippobox
\def\CMax{\bar{C}}

\def\stTD#1#2{\hbox to 0em{\mathsurround=0em %
$\stackrel{#1}{#2}$\hss} \phantom{#2}}
\def\stscript#1#2{\hbox to 0em{\mathsurround=0em %
${\scriptstyle\stackrel{#1}{#2}}$\hss} \phantom{#2}}
\def\stscriptscript#1#2{\hbox to 0em{\mathsurround=0em %
${\scriptscriptstyle\stackrel{#1}{#2}}$\hss} \phantom{#2}}
\def\st#1#2{\mathchoice{\stTD{#1}{#2}}{\stTD{#1}{#2}}%
{\stscript{#1}{#2}}{\stscriptscript{#1}{#2}}}

\def\bt{\begin{tabular}}
\def\bwt{\begin{tabular}{lcl}}
\def\et{\end{tabular}}
\def\bd{\bar{d}}
\def\bD{\underline{D}}
\def\bB{\underline{B}}
\def\bE{\underline{E}}
\def\bH{\underline{H}}
\def\tbD#1{\st{#1}{\tilde{\bf D}}}
\def\tbB#1{\st{#1}{\tilde{\bf B}}}

\def\Ssoly#1{\st{#1}{y}}
\def\rn#1{\st{#1}{r}}
\def\dn#1{\st{#1}{d}}
\def\bn#1{\st{#1}{b}}

\def\uz{\underline{z}}

\def\br{\bar{r}}

\def\cN{{\cal N}}
\def\cM{{\cal M}}
\def\clD#1{\st{#1}{{\cal D}}}
\def\cclD{{\cal D}}
\def\tclD#1{\st{#1}{\tilde{{\cal D}}}}
\def\tOm#1{\stackrel{#1}{\tilde{\Omega}}}

\def\bfgr#1{\mbox{{\boldmath $#1$}}}
\def\jm{\jmath}
\def\bjm{\bar{\jmath}}
\def\bfjm{{\bfgr \jmath}}
\def\bfbjm{\bar{\bfgr \jmath}}
\def\bfs{\bfgr{\sigma}}
\def\bs{\bar{\sigma}}
\def\ts{\tilde{\sigma}}

\def\bfbs{\bar{{\bfgr\sigma}}}
\def\dbs{\dot{\bar{\sigma}}}
\def\bfdbs{\dot{\bar{{\bfgr\sigma}}}}

\def\d{{\rm d}}
\def\x{{\bf x}}
\def\p{\partial}

\def\dfrac#1#2{{\displaystyle\frac{#1}{#2}}}
\def\dint#1#2{{\displaystyle\int\limits^{#1}_{#2}}}
\def\dsum#1#2{{\displaystyle\sum\limits^{#1}_{#2}}}

\def\J{{\cal J}}
\def\oI{{\overline {\cal I}}}
\def\oJ{{\overline {\cal J}}}

\def\mdiv{{\rm div}}
\def\rot{{\rm rot}}
\def\E{{\bf E}}
\def\B{{\bf B}}
\def\D{{\bf D}}
\def\H{{\bf H}}
\def\L{{\cal L}}
\def\oL{{\overline {\cal L}}}
\def\cR{{\cal R}}
\def\Om#1{\st{#1}{\Omega}}

\def\bfV#1{\st{#1}{\bf V}}
\def\bV2#1{\st{#1}{\bf V}^2}
\def\a#1{\st{#1}{a}}
\def\bfa#1{\st{#1}{\bf a}}
\def\sL#1{\st{#1}{L}}

\def\m#1{\st{#1}{m}}

\def\const{{\rm const}}
\def\sts#1{\st{#1}{\sigma}}
\def\bfsts#1{\st{#1}{{\bfgr\sigma}}}
\def\T{{\overline T}}

\def\m0{0}
\def\ua{\bar{a}}
\def\ab{a_\prime}
\def\ae{a_{\prime\prime}}
\def\Ve{V_{\prime\prime}}

\def\cE{{\cal E}}
\def\bcE{\bar{\cal E}}
\def\cH{{\cal H}}

\def\oA{\bar{A}}
\def\obA{\bar{\bf A}}
\def\uA{\underline{A}}
\def\ubA{\underline{{\bf A}}}
\def\uoA{\underline{\bar{A}}}
\def\uobA{\underline{\bar{\bf A}}}
\def\bA{{\bf A}}
\def\uL{\underline{{\cal L}}}

\def\bphi{{\bfgr \phi}}

\def\cP{{\cal P}}
\def\cbP{{\bfgr{\cal  P}}}
\def\sqv{\sqrt{1 {}-{} {\bf V}^2}}

\def\oLo{\bar{L}}
\def\soL#1{\st{#1}{\oLo}}

\def\uS{\underline{S}}

\def\lb#1{\label{#1} }
\newcommand{\non}{\nonumber}
\newcommand{\bref}{\begin{thebibliography}{99}}
\newcommand{\eref}{\end{thebibliography}}

\def\i{{\rm i}}

\def\bnabla{{\bfgr\nabla}}

\def\bfP#1{\st{#1}{\bf P}}

\def\eqna#1#2{\lefteqn{\mbox{\hskip #1}({\rm #2})}}
\def\aref#1{{\rm #1}}

\title{Dyons and Interactions\\[5pt] in Nonlinear (Born-Infeld) Electrodynamics}

\author{by Alexander A. Chernitskii\\
	St.Petersburg Electrotechnical University,\\
	Prof. Popov str. 5, St.Petersburg 197376,  Russia\\
	E-mail: aa@cher.etu.spb.ru}
\preprint{\jhep{12}{1999}{010}\\ \hepth{9911093}}

\abstract{
Born-Infeld nonlinear electrodynamics with point singularities having
both electric and magnetic charges are considered. Problem of
interaction between the associated soliton dyon solutions is investigated.
For the case of long-range interaction at first order by a small field
of distant solitons we obtain that the generalized Lorentz force is
acted on a dyon under consideration. Short-range interaction between
two dyons
having identical electric and opposite magnetic charges is investigated
for an initial approximation.
We consider the case when the velocities of the dyons have equal modules and
opposite directions on a common line.
It is shown that the associated field configuration has
a constant full angular momentum which
is independent of the interdyonic distance and their speed.
This property permits a consideration of
this bidyon configuration
as an electromagnetic model of charged particle with spin.
We numerically
investigate movement of the dyons in this configuration
for the case when the full electric charge equals the electron charge and
the full angular momentum equals the electron spin.
It is shown that for this case
the absolute value of relation between electric and magnetic charges
of the dyons equals the fine structure constant.
The calculation gives that the bidyon
may behave as nonlinear oscillator.
Associated
dependence of frequency on the full energy
is obtained for the initial approximation.
In the limits of the electrodynamic model we obtain that
the quick-oscillating wave packet may behave
like massive gravitating particle when it move in high background field.
We discuss the possible electrodynamic world with the oscillating bidyons
as particles.}

\begin{document}
\mathsurround=2pt

\section{Introduction}

Solitons in nonlinear electrodynamics may behave like
real particles. This analogy with classical particles is
manifested when we consider a long-range interaction of solitons
by a perturbation method \cite{I1992,I1995GR14}.
In this case a soliton is subjected to the Lorentz force
in the first order by a field of distant solitons.
The second order may give the soliton's trajectory in the form of
geodesic line for some effective Riemann space with metric
depending
on the field of distant solitons. Thus we have also
an analogy with gravitating particles. Moreover, light beams distortion
under the action of some given field
may appear as gravitational distortion
\cite{I1998JHEP}. These properties (in particular) provoke interest
to nonlinear electrodynamic models in the context of possible
unifying description for the real material world. In the framework
of this approach it may hope to solve the problem of unification
electromagnetism and gravitation.

Here we shall consider an interaction of singular
solitons in Born-Infeld nonlinear electrodynamics\footnote{It will be
recalled that
M.~Born and L.~Infeld had been considered in their article \cite{BornInfeld}
the electrodynamic model which follows from an action that had been
proposed by A.S.~Eddington in his book \cite{Eddington1924}.}.
The investigation of this nonlinear electrodynamic model with
point singularities of field is presented in the article \cite{I1998HPA},
where we have considered the singularities with electric charge.
But the model's system of equations for electromagnetic field
has an exact solution which has non-zero divergence for both
fields of electric $\D$ and magnetic $\B$ inductions at the singular point.
That is, this point for the solution has both electric and magnetic charges.
The particle with both electric and magnetic charges had been named
by J.~Schwinger as dyon (see \cite{Schwinger}).
Because the singular solution looks like
point charged particle \cite{I1998HPA}, we name this field
configuration as dyon.
In the present article we consider an interaction between these dyons
and we consider some quick-oscillating two-dyons field configuration. We also
consider like-gravitational interaction of a quick-oscillating
wave packet with given field.

\section{Basic relation for field}

Let us state the basic relation for the Born-Infeld nonlinear electrodynamics
with singularities \cite{I1998HPA} generalized to the case
with both electric and magnetic charges.
We shall use an inertial coordinate system in Minkowskian space:
$\{x^\mu\}$ (the Greek indexes take values $0,1,2,3$).
That is, components of metric is independent of time $x^0$ and
$|g_{00}| {}={} 1$, $g_{0i} {}={} 0$ (the Latin indexes take values $1,2,3$).
In this case
(see also \cite{I1998HPA})
we can write the following nonlinear Maxwell system of equations:
\begin{equation}
\label{Eq:Maxwell}
\left\{
\begin{array}{rl}
\left.
\begin{array}{rcll}
\mdiv \B  &=& 4\pi\,\bjm^0  & \eqna{22ex}{{}^\prime}
\\[2.5mm]
\mdiv \D  &=& 4\pi\,\jm^0   & \eqna{22ex}{{}^{\prime\prime}}
\end{array}\mbox{\hskip 2ex}
\right\}\;, & \eqna{19ex}{a}
\\[4ex]
\left.
\begin{array}{rcll}
\p_0 \,\B  {}+{}  \rot\E  &=& -4\pi\,\bfbjm
& \eqna{22ex}{{}^{\prime}}
\\[2.5mm]
\p_0 \,\D  {}-{}  \rot\H  &=& -4\pi\,\bfjm
& \eqna{22ex}{{}^{\prime\prime}}
\end{array}\mbox{\hskip 2ex}
\right\}\;, & \eqna{19ex}{b}
\end{array}
\right.
\end{equation}
where
\vskip-5ex
\begin{eqnarray}
&&\qquad
\begin{array}{l}
\left\{
\begin{array}{ccl}
\D &=& \dfrac{1}{\L}\left(\E  {}+{}  \alpha^2\,\J \B\right)\\[1.5ex]
\H &=& \dfrac{1}{\L}\left(\B  {}-{}  \alpha^2\,\J \E\right)
\end{array}
\right.\quad,\qquad
\left\{
\begin{array}{ccl}
\E &=& \dfrac{1}{\oL}\left(\D  {}-{}  \alpha^2\,\J \H\right)\\[1.5ex]
\B &=& \dfrac{1}{\oL}\left(\H  {}+{}  \alpha^2\,\J \D\right)
\end{array}
\right.
\quad,\\[5ex]
\begin{array}{rclcrclcrcll}
{\cal L} &=&
\sqrt{\vphantom{{\overline{\cal J}}^2}|\,1 {}-{}  \alpha^2\,{\cal I}
{}-{}  \alpha^4\,{\cal J}^2\,|}
&\;,\quad&
{\cal I} &=& {\bf E}^2 {}-{} {\bf B}^2
&\;,\quad&
{\cal J} &=& {\bf E} {}\cdot{} {\bf B}
&\;,\\[1ex]
\oL &\equiv&
\sqrt{|1 {}-{} \alpha^2 \,\oI {}-{} \alpha^4 \,\oJ^2 |}
&\;,\quad&
 \oI &=& {\H}^2 {}-{} {\D}^2
&\;,\quad&
 \oJ &=& {\H} {}\cdot{} {\D}
&\;,
\end{array}
\end{array}
\label{MatEq}
\end{eqnarray}
$\jm^\mu$ and $\bjm^\mu$ are components of electric and magnetic
singular currents such that
\begin{equation}
\lb{Def:Current}
\begin{array}{rclcrcl}
\jm^0 &\equiv&
\frac{1}{\sqrt{|g|}}\,{\displaystyle
\sum\limits_{n {}={} 1}^N}\,\dn{n}\,
\delta(\x {}-{} \bfa{n}) &\quad,\qquad&
\bfjm &\equiv&
\frac{1}{\sqrt{|g|}}\,{\displaystyle
\sum\limits_{n {}={} 1}^N}\,\dn{n}\,\bfV{n}\, \delta(\x {}-{} \bfa{n})
\quad,
\\[1em]
\bjm^0 &\equiv&
\frac{1}{\sqrt{|g|}}\,{\displaystyle
\sum\limits_{n {}={} 1}^N}\,\bn{n}\,
\delta(\x {}-{} \bfa{n}) &\quad,\qquad&
\bfbjm &\equiv&
\frac{1}{\sqrt{|g|}}\,{\displaystyle
\sum\limits_{n {}={} 1}^N}\,\bn{n}\,\bfV{n}\, \delta(\x {}-{} \bfa{n})
\quad,
\end{array}
\end{equation}
$\dn{n}$ and $\bn{n}$ are electric
 and magnetic charges of $n$-th
singularity,\\
$\bfa{n} {}={} \bfa{n}(x^0)$ is a trajectory of it and
$\bfV{n} {}\equiv{} \dfrac{\d \bfa{n}}{\d x^0}$.\\[0.5em]
Here we use the definition for the three-dimensional $\delta$-function
which is suitable for discontinuous functions $f(\x)$:
\begin{eqnarray}
\int\limits_{\Om{n}} f(\x)\,\delta(\x {}-{} \bfa{n})\,(\d x)^3  {}\equiv{}
\lim_{{\sts{n}} \to 0}\,\left\{\frac{1}{|\sts{n}|}\,
\int\limits_{{\sts{n}}}
f(\x)\;\d{{\sts{n}}} \right\} {}\equiv{}
\left\langle\vphantom{\int} f(\x) \right\rangle_n
\;,\;\;|\sts{n}|  {}\equiv{}  \int\limits_{{\sts{n}}}\d{{\sts{n}}}
\;.\;\;\;\;\;
\label{Def:delta}
\end{eqnarray}
where $\Om{n}$ is a region of three-dimensional space including
the point $\x  {}={}  \bfa{n}$,\\
 $\sts{n}$ is a closed surface enclosing this point,
$\d{{\sts {n}}} $ is an area element of the surface $\sts{n}$,\\
$|\sts{n}|$ is an area of the whole surface $\sts{n}$.

Let us define the following functions:
\begin{equation}
\lb{Def:cH}
\begin{array}{rclcrcl}
\cH &\equiv& \L {}+{} \alpha^2\,\D {}\cdot{} \E &=&
\oL {}+{} \alpha^2\,\B {}\cdot{} \H &=&
\alpha^2\,T^{00} {}+{} 1
\quad,
\\
\cbP &\equiv&
\left( \D {}\times{} \B \right)&,&
  \cP^i &=& T^{0i}
\quad,
\end{array}
\end{equation}
where $T^{\mu\nu}$ are components of symmetrical energy-momentum tensor\\
\phantom{where} (see \cite{I1998HPA} where we designated it as $\T^{\mu\nu}$).

Then, from relations (\ref{MatEq}) we can obtain the following relation
(see also \cite{BialynickiBirula,Gibbonshepth9506035}):
\begin{eqnarray}
\lb{Expr:EHDB}
& &\left\{
\begin{array}{rclclcl}
\E &=& \dfrac{1}{\alpha^2}\,\dfrac{\p \cH}{\p \D}
&=&  \dfrac{1}{\cH}\left(
\D {}-{} \alpha^2\,\cbP {}\times{} \B
\right)
&=&
\dfrac{1}{\cH}\left[\left(1 {}+{} \alpha^2\,\B^2 \right)
\D {}-{} \alpha^2\,\left(\D {}\cdot{} \B \right) \B\right]
\\[3ex]
\H &=& \dfrac{1}{\alpha^2}\,\dfrac{\p \cH}{\p \B}
&=&  \dfrac{1}{\cH}\left(
\B {}+{} \alpha^2\,\cbP {}\times{} \D
\right)
&=&
\dfrac{1}{\cH}\left[\left(1 {}+{} \alpha^2\,\D^2 \right)
\B {}-{} \alpha^2\,\left(\D {}\cdot{} \B \right) \D\right]
\end{array}
\right.
,\qquad\;
\end{eqnarray}
where\qquad\quad
$\;\;\,\cH {}={} \sqrt{1 {}+{} \alpha^2 \left(\D^2 {}+{} \B^2 \right)
{}+{} \alpha^4\,\cbP^2} {}={}$
\begin{eqnarray}
\lb{Def:cHs}
&=& \sqrt{\left(1 {}+{} \alpha^2\,\D^2 \right)
\left(1 {}+{} \alpha^2\,\B^2 \right) {}-{}
\alpha^4\,\left(\D {}\cdot{} \B \right)^2}
\quad.\qquad
\end{eqnarray}

Using relations (\ref{Expr:EHDB}) we can consider system
(\ref{Eq:Maxwell})
as the system of equations for fields $\D,\B$. This representation
is best suitable for investigation of the singular dyon solutions.

Let us introduce the two electromagnetic potentials $A_\mu$, $\oA_\mu$.
In our case the potentials have singular line for each singular charged point
(in three-dimensional space).
We must exclude such line from space
when consider any ordinary differential field model. The alternative way is
connected with using distributions or generalized functions.
As described by P.A.M.~Dirac for monopoles \cite{Dirac1948},
in this case we must include some distributions
into definitions of the potentials through derivatives.
But here we shall adhere to the first way associated with exclusion of
the singular lines. Note that the singular currents into equations
(\ref{Eq:Maxwell}) set boundary conditions at
the singular points (see \cite{I1998HPA}). In contrast,
we shall take natural boundary conditions at the singular lines
outside of the singular points.

Thus we define the potentials outside of any singular
set with help the following formulas:
\begin{equation}
\lb{Def:Potentials}
\begin{array}{rclcrcl}
F_{\mu\nu} &=& \p_\mu A_\nu  {}-{} \p_\nu A_\mu
&\quad,\qquad&
f^{\mu\nu} &=&
-\varepsilon^{\mu\nu\sigma\rho}\,\p_\sigma \oA_\rho
\end{array}
\quad,
\end{equation}
where
$\varepsilon_{0123} {}={} |g|^{1/2}$,
$\varepsilon^{0123} {}={} -|g|^{-1/2}$,
 and
\begin{equation}
\left\{
\begin{array}{rcl}
E_i&\equiv& F_{i0}\\
[0.5ex]
B^i&\equiv&
\dfrac{1}{2}\,
\varepsilon^{ijk}\, F_{jk}\\
[1.1ex]
F_{ij} &=& \varepsilon_{ijk}\, B^k
\end{array}
\right.
\qquad,\qquad
\left\{
\begin{array}{rcl}
D^i&\equiv&  f^{0i}\\
[0.5ex]
H_i&\equiv&
\dfrac{1}{2}\,\varepsilon_{ijk}\, f^{jk}\\
[1.1ex]
f^{ij} &=& \varepsilon^{ijk}\, H_k
\end{array}
\right.
\quad,
\label{Def:EBDH}
\end{equation}
where
$\varepsilon_{123} {}={} |g|^{1/2}$,
$\varepsilon^{123} {}={} |g|^{-1/2}$,
($|g_{00}| {}={} 1$, $g_{0i} {}={} 0$).

We have the following definition in three-dimensional designations:
\begin{equation}
\lb{DHBEfromAA}
\begin{array}{ll}
\left\{
\begin{array}{rcl}
\B &=& \phantom{-}\bnabla {}\times{}  \bA
\\[0.5ex]
\D &=& \phantom{-}\bnabla {}\times{}  \obA
\end{array}
\right.&\quad,\eqna{8.5em}{a}
\\[2.5ex]
\left\{
\begin{array}{rcl}
\H &=& -\bnabla \oA_0 {}+{} \p_0 \obA
\\[0.5ex]
\E &=& \phantom{-}\bnabla A_0 {}-{} \p_0 \bA
\end{array}
\right.
&\quad.\eqna{8.5em}{b}
\end{array}
\end{equation}

From the basic equation of the model outside of the singularities \cite{I1998HPA}
\begin{equation}
\label{EulerBI}
\frac{1}{\sqrt{|g|}}\,\frac{\p}{\p x^\mu}\,\sqrt{|g|}\; f^{\mu\nu} {}={} 0
\quad,\qquad
f^{\mu\nu} {}={}
\frac{1}{\L}\,\left(F^{\mu\nu}  {}-{}
\frac{\alpha^2}{2}\,\J\,
\varepsilon^{\mu\nu\sigma\rho}\,F_{\sigma\rho}\right)
\end{equation}
we can easy obtain also the following equation for the potential $A_\mu$
outside of the singular set:
\begin{equation}
\lb{Eq:EulerBIA}
C^{\mu\nu\sigma\rho}\,%
\frac{\p^2 A_\rho}{\p x^\mu \p x^\sigma} {}+{}%
\frac{\L}{2}\,f^{\mu\nu}\,\frac{\p \ln |g|}{\p x^\mu} {}={} 0%
\quad,
\end{equation}
where
\begin{eqnarray}
\lb{Def:C}
C^{\mu\nu\sigma\rho} &=&
g^{\mu\sigma}\,g^{\nu\rho} {}-{}
g^{\mu\rho}\,g^{\nu\sigma} {}-{} \alpha^2
\left( {\cal F}^{\mu\nu}\,{\cal F}^{\sigma\rho}
 {}+{} f^{\mu\nu}\,f^{\sigma\rho}
\right)
\quad,\\[2ex]
{\cal F}^{\mu\nu} &\equiv& -\dfrac{1}{2}\,
\varepsilon^{\mu\nu\alpha\rho}\,F_{\alpha\rho}
\quad.
\end{eqnarray}

\section{Boundary conditions at the singular points}

Let us integrate system of equations (\ref{Eq:Maxwell})
over a small
four-dimensional space region including only $n$-th singular point.
Using the partial integration in four-dimensional space we
obtain the following conditions:
\begin{equation}
\lb{Cond:Boundinv}
\begin{array}{rclcrcl}
\dint{\infty}{-\infty}
\Biggl[\;\dint{}{{\sts{n}}} f^{\mu\nu}\,\d {\sts{n}_\nu}
{}-{}
4\pi
\dint{}{{\Om{n}}}\jmath^\mu\,\d\Om{}
\Biggr]\,\d x^0 &=& 0
&\;,\;\;&
\dint{\infty}{-\infty}
\Biggl[\;\dint{}{{\sts{n}}}
{\cal F}^{\mu\nu}\,\d {\sts{n}_\nu}
{}-{}
4\pi
\dint{}{{\Om{n}}}\bar{\jmath}^\mu\,\d\Om{}
\Biggr]\,\d x^0 &=& 0
\;\;,
\end{array}
\quad
\end{equation}
\bt{ll}
where &
$\Om{n}$ is a three-dimensional region
including only $n$-th singular point,
\\
& $\d\st{n}{\bfs}$  is external directed element
of the closed (two-dimensional) surface\\
&  $\sts{n}$
enclosing the singular point (see also \cite{I1998HPA}),
\\
& $\d\Om{} {}\equiv{} \sqrt{|g|}\,(\d x)^3$
\quad,\qquad
$\d \sts{n}_0 {}\equiv{}
-%
\bfV{n} {}\cdot{} \d\st{n}{\bfs}$
\quad,
\\
& $\d\sts{n}_\mu\,\d x^0 {}={}
-%
\dfrac{1}{6}\,
\varepsilon_{\mu\nu\sigma\rho}\,\d x^\nu {}\wedge{}
\d x^\sigma {}\wedge{} \d x^\rho$ are components of four-vector.
\et\\
The surface $\sts{n}$ is rigidly coupled with $n$-th singular point and
move with it together.

Relations (\ref{Cond:Boundinv}) will be satisfied if we have
in three-dimensional designations
\begin{eqnarray}
\lb{Cond:Bound}
\left\{
\begin{array}{rl}
\left.
\begin{array}{rcl}
\displaystyle\int\limits_{{\sts{n}}}
\B {}\cdot{} \d\st{n}{\bfs} &=& 4 \pi\,\bn{n}
\\[2.5mm]
\displaystyle\int\limits_{{\sts{n}}}
\D {}\cdot{} \d\st{n}{\bfs} &=& 4 \pi\,\dn{n}
\end{array}\quad
\right\}
\quad,& \eqna{11ex}{a}
\\[20pt]
\left.
\begin{array}{rcl}
\displaystyle\int\limits_{{\sts{n}}}
\left[
\B \left( \bfV{n} {}\cdot{} \d\st{n}{\bfs}\right) {}+{}
\E {}\times{} \d\st{n}{\bfs}  \right] &=& 4\pi\,\bn{n}\,\bfV{n}
\\[2.5mm]
\displaystyle\int\limits_{{\sts{n}}}
\left[ \D \left( \bfV{n} {}\cdot{} \d\st{n}{\bfs}\right)
{}-{} \H {}\times{} \d\st{n}{\bfs}  \right] &=& 4\pi\,\dn{n}\,\bfV{n}
\end{array}\quad
\right\}
\quad.& \eqna{11ex}{b}
\end{array}
\right.
\end{eqnarray}
Because the surface $\sts{n}$ may be arbitrarily small,
 relations (\ref{Cond:Bound}) are boundary conditions
at the $n$-th singular point.

There is the attractive idea, proposed by A.S.~Eddington
\cite{Eddington1924},
about an invariance of theory under permutation of points of space.
We can apply this idea to the singular points. Let the model
be invariant under permutation of any two
singular points in the sense of change their charges
$\dn{n_1} {}\leftrightarrow{} \dn{n_2}\;,\;
\bn{n_1} {}\leftrightarrow{} \bn{n_2}$. Because the
theory is invariant under change of charge's sign,
in this case we have for any singular point
\begin{equation}
\dn{n} {}={} \pm \bar{d} \quad,\qquad \bn{n} {}={} \pm \bar{b} \quad,
\end{equation}
where $\bar{d}$ and $\bar{b}$ are some positive constants of space.

Thus we have the theory with three dimensional constants
$\alpha$, $\bar{d}$ and $\bar{b}$. For the suitable dimensional system we can
take $\alpha {}={} 1 $ and $\bar{d} {}={} 1$ or $\bar{b} {}={} 1$.
Thus we have the relation $\bar{d}/\bar{b}$
as the single dimensionless constant of the theory.
Below (in section \ref{Sec:Bidyon}) we shall connect
this relation with the fine structure constant.

\section{Dyon solutions and singularities}
\lb{Sec:Dsas}

System of equations (\ref{Eq:Maxwell}) with relations
(\ref{Expr:EHDB}) has the following exact dyon solution in a
cartesian coordinate system $\{y^\mu\}$:
\begin{equation}
\lb{Sol:Dyon}
\begin{array}{ll}
\bD^i {}={} \dfrac{y^i\,d}{r^3}
\quad,\qquad
\bB^i {}={} \dfrac{y^i\,b}{r^3}
\quad,&\eqna{11ex}{a}
\\[1em]
\bE^i {}={} \dfrac{y^i\,d}{r\,\sqrt{\br^4 {}+{} r^4}}
\quad,\qquad
\bH^i {}={} \dfrac{y^i\,b}{r\,\sqrt{\br^4 {}+{} r^4}}
\quad,&\eqna{11ex}{b}
\\[1.5em]
\begin{array}{rclcrcl}
\uA_0 &=& d\,\phi_0(r)
&\quad,\qquad &
\ubA &=& b\,\bphi({\bf y})
\quad,
\\[0.5ex]
\uoA_0 &=& -b\,\phi_0(r)
&\quad,\qquad &
\uobA &=& d\,\bphi({\bf y})
\quad,
\end{array}
&\eqna{11ex}{c}
\end{array}
\end{equation}
where \quad $b {}={} \pm \bar{b}$\quad,\qquad $d {}={} \pm \bar{d}$
\quad,
\begin{equation}
\lb{Def:br}
r {}={} \sqrt{y^i y_i}
\quad,\qquad
\phi_0(r) {}={}
\int\limits_\infty^r
\frac{\d r^\prime}{\sqrt{\br^4 {}+{} r^{\prime 4} } }
\quad,\qquad
\br {}\equiv{}
\left[\alpha^2\left(\bar{d}^2 {}+{} \bar{b}^2\right)\right]^{\frac{1}{4}}
\quad.
\end{equation}
The vector function $\bphi({\bf y})$ may be of two types:  with
infinite singular line and with semi-infinite one.
Its components have the following forms:
\begin{eqnarray}
\lb{Sol:TwoVar}
&&
\mbox{or}\qquad
\begin{array}{ll}
\begin{array}{rclcrclcrcll}
\phi_1&=&\dfrac{1}{r}\,\dfrac{y^2\,y^3}{\rho^2}
&\;,\;\;&
\phi_2&=&-\dfrac{1}{r}\,\dfrac{y^1\,y^3}{\rho^2}
&\;,\;\;&
\phi_3&=& 0 &\;,
\end{array}
&\eqna{1.3em}{a}
\\[2em]
\begin{array}{rclcrclcrcll}
\phi_1&=&\dfrac{1}{r}\,\dfrac{\left(r {}+{} y^3\right) y^2}{\rho^2}
&\;,\;\;&
\phi_2&=&-\dfrac{1}{r}\,\dfrac{\left(r {}+{} y^3\right) y^1}{\rho^2}
&\;,\;\;&
\phi_3&=& 0 &\;,
\end{array}
&\eqna{1.3em}{b}
\end{array}
\qquad
\end{eqnarray}
where \qquad
$\rho {\,}={\,} \sqrt{(y^1)^2 {}+{} (y^2)^2}$\quad.

Function (\ref{Sol:TwoVar}\aref{a}) has the infinite
singular line coinciding with the axis $y^3$
and functions (\ref{Sol:TwoVar}\aref{b})
has the semi-infinite singular line in positive
direction of $y^3$.
In a spherical coordinate system $\{r,\vartheta,\varphi\}$
the both functions $\bphi({\bf y})$ have non-zero
\mbox{$\varphi$-components}
only that can be written in the forms
\begin{eqnarray}
\lb{Sol:TwoVarPhy}
&&
\begin{array}{rclcrcl}
&(a)&&\qquad\quad\mbox{or}\quad\qquad&&(b)&
\\[0.7ex]
\phi_\varphi &=& -\dfrac{\cot{\vartheta}}{r}
&&
\phi_\varphi &=& -\dfrac{1}{r}\,\cot{\dfrac{\vartheta}{2}}
\end{array}
\end{eqnarray}
accordingly.

At the singular line the potentials $\bA,\obA$ are devoid of
defined direction and their absolute values are infinity. At the singular point
$r {}={} 0$ the vectors $\E,\H,\D,\B$ are devoid of
defined direction and absolute values of $\D,\B$ are infinity.

The full electromagnetic energy
(see (\ref{Def:EnergyMomentum}) below)
of dyon solution (\ref{Sol:Dyon}) is
\begin{eqnarray}
\lb{Def:bcE}
\bcE&=&
\frac{2}{3}\left(\alpha^{-2}\,\beta\,\br^3\right)\quad,\\[0.5ex]
&&\mbox{where}\quad
\beta {}\equiv{} \int\limits_0^\infty\frac{\d r}{\sqrt{1 {}+{} r^4}} {}={}
\frac{\left[\Gamma (\frac{1}{4})\right]^2}{4\,\sqrt{\pi}}
\approx 1.8541
\quad.
\non
\end{eqnarray}

The dyon's energy has a space localization region. Let us denote
the sphere-enclosed dyon's energy as ${\cE}^\prime$ for radius of
the sphere $r^\prime$.
Center of the sphere is at the origin of the coordinates $\{y^i\}$.
The numerical calculation gives that
${\cE}^\prime {}={} 0.5\,\bar{\cE}$ for $r^\prime {}={}\br $
and
${\cE}^\prime {}={} 0.95\,\bar{\cE}$ for $r^\prime {}={} 10\,\br $.

With help of Lorentz transformation, shift, and rotation of the
coordinate system $\{y^i\}$ we can obtain the following
moving dyon solution in the coordinates $\{x^\mu\}$:
\begin{eqnarray}
&&
\lb{Sol:Dyonm}
\begin{array}{ll}
\left\{
\begin{array}{rcl}
D^i &=&
\left.\left( \oLo^i_j\,\bD^j {}-{}
\varepsilon^{ijl}\,V_j\,\bH_l \right)\right/\sqv\\[7pt]
B^i &=&
\left.\left( \oLo^i_j\,\bB^j {}+{}
\varepsilon^{ijl}\,V_j\,\bE_l \right)\right/\sqv
\end{array}
\right.
&\quad,\eqna{5em}{a}\\[20pt]
\left\{
\begin{array}{rcl}
E^i &=&
\left.\left( \oLo^i_j\,\bE^j {}-{}
\varepsilon^{ijl}\,V_j\,\bB_l \right)\right/\sqv\\[7pt]
H^i &=&
\left.\left( \oLo^i_j\,\bH^j {}+{}
\varepsilon^{ijl}\,V_j\,\bD_l \right)\right/\sqv
\end{array}
\right.
&\quad,\eqna{5em}{b}\\[20pt]
\left\{
\begin{array}{rcl}
A_\mu &=& L^0_{.\mu}\,\uA_0 {}+{}
L^i_{.\mu}\,\cR^j_{.i}\,\uA_j({\bf z})\\[0.5ex]
\oA_\mu &=& L^0_{.\mu}\,\uoA_0 {}+{}
L^i_{.\mu}\,\cR^j_{.i}\,\uoA_j({\bf z})
\end{array}
\right.
&\quad,\eqna{5em}{c}\\[20pt]
\end{array}
\;\;
\end{eqnarray}
where $\bD^i,\bB^i,\bE^i,\bH^i,\uA_\nu,\uoA_\nu$
are defined by formulas (\ref{Sol:Dyon})\\
with $y^i {}={} L^i_j \left(x^j {}-{} V^j\,x^0 {}-{} a_0^j\right)$
and $z^i {}={} \cR^i_{.j}\,y^j\;\;$,\\
$L^\nu_{.\mu}$ is the Lorentz transformation matrix,
$L^i_j\,\oLo^j_l {}={} \delta^i_l\;\;$,
\begin{equation}
\label{Def:Lorenz}
\begin{array}{l}
\begin{array}{rclcrcrcl}
L^0_{.0}  &=& \dfrac{1}{\sqrt{1 {}-{} \bV2{}}}
&\;\;,\quad&
L^0_{.i}  &=&  L_{i0}  &=&
 -\dfrac{V_i}{\sqrt{1 {}-{} \bV2{}}}
\end{array}
\quad,
\\[3ex]
\begin{array}{rclcrcl}
L^i_j &=& \delta^i_j {}+{}
\left(\dfrac{1}{\sqrt{1 {}-{} \bV2{}}}  {}-{}  1\right)\,
\dfrac{V^i V_j}{{\bf V}^2}
&\;\;,\quad&
\oLo^i_j &=& \delta^i_j {}+{}
\left({
\sqrt{1 {}-{} \bV2{}} }   {}-{}  1\right)\,
\dfrac{V^i V_j}{{\bf V}^2}
\end{array}
\;\;,
\end{array}
\end{equation}
$a_0^j$ are components of an initial position,
$\cR^i_{.j}$ is a rotation matrix.

\section{Variational principle for two potentials}

Here we propose some action such that the associated variational
principle gives system of equations (\ref{Eq:Maxwell})
with relations (\ref{Expr:EHDB}) and
boundary conditions (\ref{Cond:Bound}).
This action has the following form:
\begin{eqnarray}
\lb{Def:ActDual}
\uS &=& \int \left[ \left(\uL {}+{} 1
\right)
 {}+{} 2\pi\,\alpha^2
\left(\jm^\mu\,A_\mu {}-{} \bjm^\mu\,\oA_\mu   \right)\right]
\sqrt{g}\, (\d x)^4
\quad,
\\
\lb{Expr:uLbasic}
\mbox{where }&\;\;&
\uL {\,}\equiv{\,}  -\cH {}+{} \frac{\alpha^2}{2}\left(
\E {}\cdot{} \D {}+{} \H {}\cdot{} \B \right)
\quad,
\\
\non
&&\cH {\,}={\,} \cH (\D,\B) \mbox{ according to (\ref{Def:cHs}),}
\\
\non
&&\mbox{$\E,\B$ are represented by $A_\mu$ and
$\D,\H$ are represented by $\oA_\mu$ with (\ref{DHBEfromAA}).}
\end{eqnarray}

Using definition (\ref{Def:cH}) we can easy write also the following
expression for $\uL$:
\begin{eqnarray}
\lb{Expr:uLm}
\uL  &=& -\dfrac{1}{2}\left(\L {}+{} \oL\right)
\quad.
\end{eqnarray}
Thus the action $\uS$ is invariant under general transformation
of coordinates.

If we substitute expression (\ref{Expr:uLm}) for $\uL$
into the action $\uS$ and take into account that
$\L {}={} \L(\p_\mu A_\nu),\,\oL {}={} \oL(\p_\mu \oA_\nu)$,
then independent formal variation of eight components of the potentials
$A_\mu,\,\oA_\nu$ for the principle $\delta \uS {}={} 0$ gives
the following eight equations:
\begin{equation}
\lb{Eq:TwoPot}
\dfrac{1}{\scriptstyle{\sqrt{|g|}}}\,\dfrac{\p}{\p x^\mu}
\left[\scriptstyle{\sqrt{|g|}}\,
\dfrac{\p \L}{\p (\p_\nu A_\mu)}\right] {}={} 4\pi\,\jmath^\nu
\;,\quad
\dfrac{1}{\scriptstyle{\sqrt{|g|}}}\,\dfrac{\p}{\p x^\mu}
\left[\scriptstyle{\sqrt{|g|}}\,
\dfrac{\p \oL}{\p (\p_\nu \oA_\mu)}\right] {}={} -4\pi\,\bar{\jmath}^\nu
\;.\;\;
\end{equation}
For simplicity we didn't consider here the singular
lines of the potentials.

Formally we can consider system of equations
(\ref{Eq:Maxwell}) with relations (\ref{MatEq}) as the system
of eight equations for the eight unknown functions $A_\mu,\,\oA_\nu$
{\mathsurround=0ex
((\ref{Eq:Maxwell}\aref{$a^\prime$}),(\ref{Eq:Maxwell}\aref{$b^\prime$})
for $\oA_\mu$ and
(\ref{Eq:Maxwell}\aref{$a^{\prime\prime}$}),(\ref{Eq:Maxwell}\aref{$b^{\prime\prime}$})
for $A_\mu$)}. Such system
of equations agree with system (\ref{Eq:TwoPot}).

Now let us consider the action $\uS$ (\ref{Def:ActDual}) with
the function $\uL$ in form (\ref{Expr:uLbasic}).
In this case the dependence of Lagrangian
on the derivatives
of potentials differs from one that we have for the function $\uL$ in
form (\ref{Expr:uLm}). Thus the appropriate variational principles
with the function $\uL$ in forms (\ref{Expr:uLbasic})
and (\ref{Expr:uLm}) are different.

Substituting definitions (\ref{DHBEfromAA}), (\ref{Def:Current}),
(\ref{Def:delta}) into (\ref{Def:ActDual})
we obtain the following expression for the action that doesn't
contain $\delta$-functions:
\begin{eqnarray}
\lb{Def:ActDualm}
\uS &=&
\int\left(\uL {}+{} 1\right) \sqrt{|g|}\, (\d x)^4
{}+{}
\\
&&
{}+{} 2\pi\,\alpha^2\,
\sum\limits_{n {}={} 1}^N \int\left\langle
\dn{n}\left(\vphantom{\obA} A_0 {}+{}
\bfV{n} {}\cdot{} \vphantom{\obA}\bA\right)
{}-{}
\bn{n}\left(\oA_0 {}+{} \bfV{n} {}\cdot{} \obA \right)
 \right\rangle_{\!\!n} \d x^0
\quad,
\non\\[0.5ex]
\uL &=& -\cH {}+{}
\frac{\alpha^2}{2}\left[
\left(\vphantom{\obA}\bnabla A_0 {}-{} \p_0 \bA \right)
{}\cdot{} \left(\bnabla {}\times{} \obA\right) {}+{}
\left(\p_0 \obA {}-{} \bnabla \oA_0 \right)
{}\cdot{}
\left(\vphantom{\obA}\bnabla {}\times{} \bA\right)
 \right]
\;\;.\qquad
\lb{Def:uLfromAA}
\end{eqnarray}
\FIGURE{
\lb{Fig:surface}
\begin{picture}(160,130)
\put(-20,0){
\begin{picture}(160,130)
\put(60,60){\circle*{3}}
\put(60,60){\oval(50,50)[b]}
\put(60,60){\oval(50,50)[tl]}
\put(21,55){$\sts{n}$}
\put(70,110){$\dbs$}
\put(60,60){\line(1,1){70}}
\put(60,85){\line(1,1){50}}
\put(85,60){\line(1,1){60}}
\put(110,85){\vector(1,-1){15}}
\put(110,56){$\d\bfbs {}={} \d\bfdbs$}
\put(79,41){\vector(1,-1){15}}
\put(79,13){$\d\bfbs {}={} \d\bfsts{n}$}
\end{picture}
}
\end{picture}
\caption{Part of the surface $\bs$ for the dyon field configuration with
semi-infinite singular line.}
}
{\hfuzz=4pt
Note, though potentials $A_\mu$, $\oA_\mu$ (\ref{Sol:Dyonm}\aref{c})
for the dyon solutions are infinite at the singular lines,
its averaging at the singular points
$\langle A_\mu \vphantom{\oA_\mu}\rangle_n$,
$\langle \oA_\mu \rangle_n$ are finite. Thus the action $\uS$
is finite for the dyon solution.

Because definition of the potentials (\ref{DHBEfromAA}) may be
used only outside of the singular set, expression for $\uL$
(\ref{Def:uLfromAA}) is true only outside of the singular lines.
Let us enclose all singular set in a multi-tuple connected
surface $\bs$ with external (relative to the singularity)
surface element $\d\bfbs$. Let this surface be composed
of parts of surfaces $\sts{n}$ for the $N$ segregated singular points
and multi-tuply connected tubular surface $\dbs$
(in three-dimensional space)  for
parts of the singular lines outside neighbourhoods of the $N$
points. For example, the surface $\bs$ for the semi-infinite
singular line can have image as it is shown in
figure \ref{Fig:surface}.

}

Let us designate the action $\uS$ in space outside of the singular set
as $\overline{\uS}$. We can obtain variation
of this action with partial integration. Thus we have
\begin{eqnarray}
\non
&&\delta\overline{\uS} {}={}
\dint{\infty}{-\infty}\d x^0\left\{
\dint{}{\overline\Omega}\left[
\left(\alpha^2\,\p_0\D  {}-{}
\bnabla {}\times{} \dfrac{\p\cH}{\p\B}\right) {}\cdot{} \delta\bA
{}-{}
\left(\alpha^2\,\p_0\B{}+{}
\bnabla {}\times{} \dfrac{\p\cH}{\p\D} \right) {}\cdot{} \delta\obA
\right]
\d\Om{}
\right.\\[1ex]
\non
&&{}-{} \dfrac{\alpha^2}{2}
\dint{}{\bs} \left[
\left(\D {}\cdot{} \d \bfbs\right) \delta A_0
{}-{} \left(\B {}\cdot{} \d \bfbs\right)\delta \oA_0 \right]  {}+{}
2\pi\,\alpha^2 \dsum{}{n}
\left(\dn{n}\,\left\langle\vphantom{\oA_0}\delta A_0\right\rangle_n  {}-{}
\bn{n}\,\left\langle\delta \oA_0\right\rangle_n \right)
\\[1ex]
\lb{Exp:deltauS}
&& {}+{} \dint{}{\bs} \left[ \dfrac{\p\cH}{\p\D} {}\times{} \d\bfbs
{}-{} \dfrac{\alpha^2}{2}\left(\E {}\times{} \d\bfbs
{}+{} \B \,\d\bs_0 \right)
\right] {}\cdot{} \delta\obA
{}-{} 2\pi\,\alpha^2\dsum{}{n} \bn{n}\,\bfV{n} {}\cdot{}
\left\langle\delta \obA\right\rangle_n
\\[1ex]
\non
&& {}+{} \dint{}{\bs} \left[ \dfrac{\p\cH}{\p\B} {}\times{} \d\bfbs
{}-{} \dfrac{\alpha^2}{2}\left(\H {}\times{} \d\bfbs
{}-{} \D \,\d\bs_0 \right)
\right] {}\cdot{} \delta\bA
{}+{} 2\pi\,\alpha^2\dsum{}{n} \dn{n}\,\bfV{n} {}\cdot{}
\left\langle\vphantom{\obA}\delta \bA\right\rangle_n
\\[1ex]
\non
&&{}+{}
\left.
2\pi\,\alpha^2\dsum{}{n}\left[\left\langle
\dn{n}\left(
\E  {}+{}
\bfV{n} {}\times{}
\B \right) {}+{} \bn{n}\left(
\H  {}-{} \bfV{n} {}\times{}
\D \right)
\right\rangle_{\!\!n} \!\cdot\delta\bfa{n}\,\right]
{}-{} \dint{}{\bs} \uL\, \d\bfbs {}\cdot{} \delta\st{\bs}{\bf a}
\right\}
\end{eqnarray}
\bt{ll}
where &$\overline\Omega$ is three-dimensional space
without the region bounded by surface $\bs$,\\
&
$\d\bs_0 {}\equiv{} -\left(\st{\bs}{\bf V} {}\cdot{} \d\bfbs\right)\;$
,
\quad
$\st{\bs}{\bf V} {}\equiv{} \dfrac{\d\st{\bs}{\bf a}}{\d x^0}\;$
,\quad
$\st{\bs}{\bf a}\;$ is a points of the surface $\bs\;$.
\et

Now we contract the surface $\dbs$ to the singular lines
outside neighbourhoods of the $N$ singular points.
Next we contract the surfaces $\sts{n}$ to the $N$ singular points.
As result we have $\delta\overline{\uS} {}\to{} \delta\uS$.

Let us find stationary conditions for the action $\uS$.
By the general principle of calculus of variations
\cite{CurantHilbert}, we can take
$\delta \bfa{n} {}={} \delta \st{\bs}{\bf a} {}={} 0$ at first.
We can also take continuous variations $\delta A_\mu,\,\delta \oA_\mu$
and make, firstly, $\delta A_\mu {}={} 0,\,\delta \oA_\mu {}={} 0$
at the singular lines.
Then, according to the first line in expression (\ref{Exp:deltauS}),
we have
\begin{equation}
\lb{Eq:Homo}
\alpha^2\,\p_0\B {}+{}
\bnabla {}\times{} \dfrac{\p\cH}{\p\D} {}={} 0
\quad,\qquad
\alpha^2\,\p_0\D {}-{}
\bnabla {}\times{} \dfrac{\p\cH}{\p\B} {}={} 0
\end{equation}
in the space outside of the singular lines.
Using definition of the fields $\B,\D,\E,\H$
through potentials (\ref{DHBEfromAA}),
we obtain from equations (\ref{Eq:Homo}) that
\begin{equation}
\lb{Expr:EHDBp}
\E {}={} \dfrac{1}{\alpha^2}\,\dfrac{\p \cH}{\p \D}
\quad,\qquad
\H {}={} \dfrac{1}{\alpha^2}\,\dfrac{\p \cH}{\p \B}
\quad.
\end{equation}

Now we take $\delta A_\mu {}\neq{} 0$ and $\delta \oA_\mu {}\neq{} 0$
at the singular lines for continuous variations
$\delta A_\mu,\,\delta \oA_\mu$. We assume that the fields $\D,\B,\E,\H$
are continuous at the singular lines outside
the segregated singular points. Then we obtain boundary conditions
(\ref{Cond:Bound}) from expression (\ref{Exp:deltauS})
(line 2-4), using formulas (\ref{Expr:EHDBp}).

Finally, we take $\delta\bfa{n} {}\neq{} 0$,
$\delta\st{\bs}{\bf a} {}\neq{} 0$.
It would appear reasonable that at the singular lines
\begin{equation}
\lb{Cond:BoundLa}
\lim\limits_{\bs {}\to{} 0} \dint{}{\bs} \uL\, \d\bfbs {}={} 0
\quad.
\end{equation}
Then, from the last line in (\ref{Exp:deltauS})
we have the following condition at the $n$-th singular point:
\begin{eqnarray}
\left\langle\dn{n}\left(
\E  {}+{}
\bfV{n} {}\times{}
\B \right) {}+{}
\bn{n}\left(
\H {}-{} \bfV{n} {}\times{}
\D \right)\right\rangle_{\!n}
&=& 0
\quad.
\label{Cond:BoundAngle}
\end{eqnarray}
This condition generalize one
for purely electrical singularity ($\bn{n} {}={} 0$), which had obtained
in the article \cite{I1998HPA}, to the case of dyon singular point.
Using definition of singular currents (\ref{Def:Current}) we
obtain also the following form of the condition:
\begin{eqnarray}
\lb{Cond:Fjfbj}
F_{\mu\nu}\,\jm^\nu {}-{}
\dfrac{1}{2}\,\varepsilon_{\mu\nu\sigma\rho}\,f^{\sigma\rho}\,\bjm^\nu
&=& 0
\quad.
\end{eqnarray}

Thus the variational principle for action $\uS$
(\ref{Def:ActDual}), (\ref{Def:ActDualm}) gives system of equations
(\ref{Eq:Maxwell}) and point boundary conditions (\ref{Cond:Bound}),
(\ref{Cond:BoundAngle}).

Notice that here we don't use a concrete form for the function $\cH (\D,\B)$.
Thus this derivation is suitable for any electrodynamic models.

\section{Conservation laws and comments about dimensions}
\lb{Sec:ConsLaw}

Using equations (\ref{Eq:Maxwell}) with formulas (\ref{Expr:EHDB})
and condition (\ref{Cond:Fjfbj})
we can check directly the following conservation law for symmetrical
energy-momentum tensor (in cartesian coordinates):
\begin{eqnarray}
\lb{ConsLaw:EMT}
\dfrac{\p T^{\mu\nu}}{\p x^\mu} &=& 0
\quad,
\end{eqnarray}
where
\begin{eqnarray}
\lb{Def:EMT}
\begin{array}{rcl}
T^{00} &=& \alpha^{-2}\left(\cH {}-{} 1 \right)
\quad,
\\[1.2ex]
T^{0i} &=& \varepsilon^{ijl}\,D_j\,B_l {\,}={\,}
\varepsilon^{ijl}\,E_j\,H_l
\quad,
\\[1.2ex]
T^{ij} &=& \delta^{ij}
\left[\D\cdot\E {}+{} \B\cdot\H {}-{}
T^{00}\right] {}-{}
\left(D^i\,E^j {}+{} B^i\,H^j\right)
\quad.
\end{array}
\end{eqnarray}

Let us introduce the following notations for full energy and momentum
in the region $\tOm{}$ of three-dimensional space $\Om{}$.
\begin{eqnarray}
\lb{Def:EnergyMomentum}
\begin{array}{rclcrcl}
{\cE}&=&\dfrac{1}{4\pi\,\alpha^2}
\dint{}{\tOm{}} \left(\cH {}-{} 1 \right)\,\d\Om{}
&\quad,\qquad&
{\bf P} &=& \dfrac{1}{4\pi}
\dint{}{\tOm{}} \cbP\,\d\Om{}
\quad,
\end{array}
\end{eqnarray}
where $\cP^i {}\equiv{} T^{0i}$ or $\cbP {}\equiv{} \D {}\times{} \B$ .

Let the region $\tOm{}$ may be moving, that is we have
$\tilde{\Omega} {}={} \tilde{\Omega}(x^0)$.
Then, by integration of conservation law (\ref{ConsLaw:EMT}) over the
region $\tOm{}$ we obtain the following integral conservation law for
energy-momentum:
\begin{eqnarray}
\lb{Cons:EM}
\begin{array}{rcll}
\dfrac{\d {\cE}}{\d x^0}
&=&-\dfrac{1}{4\pi}\,\dint{}{\ts} \cbP {}\cdot{} \d\tilde{\bfs}
{}-{}
\dfrac{1}{4\pi\,\alpha^2}\,\dint{}{\ts} \left(\cH {}-{} 1\right) \d\ts_0
\quad,&\eqna{10ex}{a}\\[3ex]
\dfrac{\d P^i}{\d x^0}
 &=& -\dfrac{1}{4\pi}\dint{}{\ts} T^{ij}\, \d\ts_j
{}-{} \dfrac{1}{4\pi}\dint{}{\ts} \cP^i\, \d\ts_0
\quad,&\eqna{10ex}{b}
\end{array}
\end{eqnarray}
where $\ts$ is the surface that bounds the region $\tOm{}$,\\
\phantom{where} $\d\ts_j$ is the external element of this surface,\\
\phantom{where} $\d\ts_0 {}\equiv{} -\tilde{\bf V} {}\cdot{} \d\tilde{\bfs}$
, \quad $\tilde{\bf V}$ is the velocity of surface's point.

\vskip1.5ex
It is evident, from (\ref{ConsLaw:EMT}) we have also the following conservation
law for angular momentum tensor:
\begin{eqnarray}
\lb{ConsLaw:AMT}
\dfrac{\p M^{\mu\nu\rho}}{\p x^\mu} &=& 0
\quad,
\end{eqnarray}
where
\begin{eqnarray}
M^{\mu\nu\rho} &=& T^{\mu\nu}\,x^\rho {}-{} T^{\mu\rho}\,x^\nu
\quad.
\end{eqnarray}

We introduce the following notation for vector of full angular momentum
in the region $\tOm{}$:
\begin{eqnarray}
\lb{Def:VectAngMom}
{\bf M} &=& \frac{1}{4\pi}\dint{}{\tOm{}} \left( \cbP {}\times{} \x \right)
\d\Om{}
\quad.
\end{eqnarray}

Let us take up somewhat the dimensions of the quantities.
Because we don't
explicitly use the speed of light $c$, the temporary coordinate $x^0$ has
a dimension identical to that for the space coordinates $x^i$,
that is dimension of length.
(But we can make substitution $x^0 {}\to{} c\,t$ for all time.)
For the same reason the momentum has a dimension identical to that for the
energy. And dimension of the full angular momentum is
$[M] {}={} [P]\,[x] {}={} [\cE]\,[x]$. In this case the dimension of mass
is identical to that for the energy.

\section{Method for investigation of interaction between dyons}
\lb{Sec:PertMethod}

We shall consider system (\ref{Eq:Maxwell}) with relations (\ref{Expr:EHDB})
as system of equations for the fields $\D,\,\B$.
Let us write this system in the following formal operational form:
\begin{equation}
\lb{Eq:Nchi}
\cN \cclD  {}={} 0
\quad,\qquad
\mbox{where}
\quad
\cclD {\,}\equiv{\,}
\left(
\begin{array}{c}
\D\\[0.25ex]
\B
\end{array}
\right)
\quad.
\end{equation}

We can search a solution of equation (\ref{Eq:Nchi}) in the form
of the following formal series:
\begin{equation}
\lb{Series}
\cclD {}={} \cclD^{(0)} {}+{} \cclD^{1)} {}+{} \cclD^{2)}
{}+{} ... \quad,\qquad
\cclD^{(k)} {}\equiv{} \cclD^{(0)} {}+{}
\sum^k_{l {}={} 1} \cclD^{l)}
\quad,
\end{equation}
\bwt
where & $\cclD^{(0)}$ & is an initial approximation,\\
& $\cclD^{ k)}$ & is $k$-th correction,\\
& $\cclD^{ (k)}$ & is $k$-th approximation.
\et

Let us expand the operator $N$ into equation (\ref{Eq:Nchi})
near to the $k$-th approximation $\cclD^{ (k)}$ and keep only linear terms
by the correction $\cclD^{k+1)}$. Thus we obtain the following equation
for the correction:
\begin{equation}
\lb{Eq:Pert1}
\left\{ \cN^\prime \left[\cclD^{(k)}\right] \right\}
\cclD^{k+1)} {}={}
- \cN \cclD^{(k)}
\quad,
\end{equation}
where $\cN^\prime$ is Frechet's derivative for the operator $\cN$.

Equation (\ref{Eq:Pert1}) gives the following iterative process,
which is called Newton's method for operators \cite{HustonPym}:
\begin{equation}
\lb{iter1}
\cclD^{(k+1)} {}={} \cclD^{(k)}  {}-{}
\left\{
\cN^\prime\left[\cclD^{(k)}\right]
\right\}^{-1}
\left[ \cN\, \cclD^{(k)}\right]
\quad.
\end{equation}

An alternative possible way is based on using the principle of contractive
mappings. Let us rewrite system (\ref{Eq:Nchi}) in the following
form:
\begin{equation}
\lb{Eq:Mchi}
\cM \cclD  {}={} \tilde{\cN} \cclD
\quad,\qquad
\mbox{where}
\quad
\tilde{\cN} \cclD {\,}\equiv{\,}
\cM \cclD {}-{} \cN \cclD
\end{equation}
and $\cM$ is the operator represented by left parts of
linear Maxwell equations for $\D {}={} \E$, $\B {}={} \H$.

In this case we can easy obtain the appropriate inverse operator
$\cM^{-1}$. As result we have the following representation for
equation (\ref{Eq:Nchi}):
\begin{equation}
\lb{Eq:Nchi1}
\cclD {}={} \cM^{-1}\,\tilde{\cN}\,\cclD
\quad.
\end{equation}
If the operator $\cM^{-1}\,\tilde{\cN}$ is contractive
(for some given class of functions including the initial approximation),
then the iterative process
\begin{equation}
\lb{iter2}
\cclD^{(k+1)} {}={} \cM^{-1}\,\tilde{\cN}\,\cclD^{(k)}
\end{equation}
converges to the solution.

The convergence of iterative processes (\ref{iter1}), (\ref{iter2})
to some solution is defined by conditions that depend
on
initial approximation \cite{HustonPym}.
But we can use these procedures, even if it doesn't converge.
In this case we may have some number of iteration for best
approximation  to the solution.
In any case the initial
approximation $\cclD^{(0)}$ must be sufficiently
close to some exact solution.

We can take the sum of some solutions as initial approximation.
Let us take moving dyons  $\clD{n}$ (\ref{Sol:Dyonm}) as these solutions.
Thus we have
\begin{equation}
\lb{Appr:Init}
\clD{}^{(0)} {}={} \sum_{n}^N \clD{n} \quad.
\end{equation}

The field configuration $\D,\,\B$ for the exact moving dyon solution
has six free parameters:
three components of the velocity and
three components of the initial position.
We can assume that these parameters for the dyons into
initial approximation (\ref{Appr:Init}) are functions of time.
This is standard practice for similar problems. But, as it can show,
if we take that the initial positions depend on time, then
boundary conditions (\ref{Cond:Bound}\aref{b}) are broken for
the dyon field configuration. Thus we may have only
three components of velocity as dyon's parameters depending on time.

We believe that each dyon has an optional trajectory
${\bf x} {}={} \bfa{n} (x^0)$ such that
\begin{equation}
\lb{Exp:Ssoly}
\Ssoly{n}^i {}={} \sL{n}^{i}_{j} \left(x^j {}-{} \a{n}^{i}\right)
\;\;\;,\qquad
\frac{\d \bfa{n}}{\d x^0} {}\equiv{} \bfV{n}
\;\;\;,\qquad
\sL{n}^{i}_{j} {}={} \sL{n}^{i}_{j} (\bfV{n})
\;\;\;,\qquad
\soL{n}^{i}_{j} {}={} \soL{n}^{i}_{j} (\bfV{n})
\;\;\;,
\qquad
\end{equation}
\bt{ll}
where&$\Ssoly{n}^i$ are included into expressions
(\ref{Sol:Dyon}\aref{a}),(\ref{Sol:Dyon}\aref{b}),\\
&$\soL{n}^{i}_{j}$ are included into expressions
(\ref{Sol:Dyonm}\aref{a}).
\et\\
The trajectories $\bfa{n}(x^0)$ must be obtained.

In the same manner as for field we can represent the functions $\bfa{n}(x^0)$
by the following series:
\begin{equation}
\lb{TrSeries}
\begin{array}{rclcrcl}
\bfa{n} &=& \bfa{n}^{(0)} {}+{} \bfa{n}^{1)} {}+{} \bfa{n}^{2)}
{}+{} ... &\quad,\qquad&
\bfa{n}^{(k)} &\equiv& \bfa{n}^{(0)} {}+{}
\dsum{k}{l {}={} 1} \bfa{n}^{l)}
\quad,
\end{array}
\end{equation}
\bwt
where & $\bfa{n}^{ (0)}$ & is the trajectory of $n$-th dyon for
the initial approximation,\\
& $\bfa{n}^{ k)}$ & is $k$-th correction to the trajectory,\\
& $\bfa{n}^{ (k)}$ & is $k$-th approximation for the trajectory.
\et.

If we have obtained the trajectories in initial
approximation $\bfa{n}^{ (0)}(x^0)$ with help some method, then we can
obtain the first correction
$\clD{}^{1)}$ as solution to equation (\ref{Eq:Pert1})
or from formula (\ref{iter2}) for $k {}={} 0$.
This correction $\clD{}^{1)}$ is a radiation of the dyons
accelerated according to their trajectories $\bfa{n}^{(0)} (x^0)$.
But the radiation of the dyons modifies their trajectories.
We describe
these modifications by functions $\bfa{n}^{1)}$ that correct the
trajectories $\bfa{n}^{(0)}$.
Thus we must obtain the corrections $\bfa{n}^{k)}$ to the
trajectories $\bfa{n}^{(k-1)}$  in any step of iteration.
In effect, this is a correction of the initial approximation
(through the correction of the trajectories) for every iteration step.

A direct calculation gives that
all boundary conditions (\ref{Cond:Bound}), (\ref{Cond:BoundAngle})
are satisfied for initial field configuration (\ref{Appr:Init})
with (\ref{Exp:Ssoly}) and optional trajectories of the dyons.
 This is distinction of this method from one that
had used in the article \cite{I1998HPA}, where we take the four components
of the potential $A_\mu$ as unknown functions.
But using the fields $\D,\,\B$ as unknown functions,
the present method is more suitable for
satisfaction to all point boundary conditions in any iteration step.

In the present article we shall use integral energy-momentum
conservation law (\ref{Cons:EM}) for obtaining trajectories
of the interacting dyons.

Notice also the following.
Because near  $n$-th singular point absolute values of
the fields $\st{n}{\D}$, $\st{n}{\B}$ tends to infinity,
for any finite given external field (of other dyons)
we can define some neighbourhood of the singular point,
where this external field may be believed to be small.
Thus the consideration of any (small) distances between
the dyons in initial approximation may be worthwhile.

\section{Interaction of dyon with a small given field}

Let us consider the interaction of moving dyon  with
a small given field $\tclD{n}^i$\break ($i {}={} 1,...,6$),
(\mbox{$\st{n}{\tilde{\E}} {}\approx{} \st{n}{\tilde{\D}}$},
\mbox{$\st{n}{\tilde{\H}} {}\approx{} \st{n}{\tilde{\B}}$})
in initial approximation. In particular, this
small field may represent the field of other dyons that are at
sufficiently long distance
from the dyon under consideration.

In this case we consider the space region $\tOm{n}$ that include
the dyon under consideration only.
The region $\tOm{n}$ move
with $n$-th singularity together.
We can take a maximal absolute value of
the functions $\tclD{n}^i$ in the region $\tOm{n}$
as the small order parameter $\epsilon$ for the problem
of finding the first approximation in the region $\tOm{n}$.
Here we consider the case when the region $\tOm{n}$ has
more size than one for dyon localization region
(see section (\ref{Sec:Dsas})).

Thus  for the region $\tOm{n}$ we take
\begin{eqnarray}
\clD{}^{(0)} &=& \clD{n} {}+{} \tclD{n}
\quad.
\end{eqnarray}

For obtaining dyon's trajectory equation we use formula
(\ref{Cons:EM}\aref{b}).
We consider the case when the field $\tclD{n}$ is
approximately constant into the region $\tOm{n}$.
Thus we have
\begin{eqnarray}
\lb{Def:Pn}
&&
\begin{array}{rcccl}
\dfrac{\d\bfP{}^{(0)}}{\d x^0} &=&
\dfrac{\d\bfP{n}}{\d x^0} &=&
\dfrac{\d}{\d x^0}\,\dfrac{\bar{\cE}\,\bfV{n}}{\sqrt{1 {}-{} \bV2{n}}}
\end{array}
\quad,
\end{eqnarray}
where $\bar{\cE}$ is defined into (\ref{Def:bcE}).
Here we take into account that size of the region $\tOm{n}$ is much more
than $\br$ (\ref{Def:br}).

Now we expand the components $T^{ij}$ as Taylor series
near the $n$-th dyon's field configuration.
We restrict the consideration to the terms of no more than
the first order by $\epsilon$. Thus we have
\begin{eqnarray}
\lb{Expand:TijFirst}
T^{ij} &=& \st{n}{T}^{ij} {}+{}
\dfrac{\p\st{n}{T}^{ij}}{\p \clD{}^l}\,\tclD{n}^l {}+{} \dots
\quad,
\end{eqnarray}
where\quad $l {}={} 1,...,6\;$\quad,\qquad
$\;\st{n}{T}^{ij} {}\equiv{}
T^{ij}\Bigr|_{\clD{}^l {}={} \clD{n}^l}\;$\quad,\qquad
$\;\dfrac{\p\st{n}{T}^{ij}}{\p \clD{}^l} {}\equiv{}
\dfrac{\p T^{ij}}{\p \clD{}^l}\Biggr|_{\clD{}^l {}={} \clD{n}^l}$\quad.

According to formulas (\ref{Def:EMT}) we have
\begin{eqnarray}
\lb{Exp:dTijdDB}
\begin{array}{rcl}
\dfrac{\p T^{ij}}{\p D^q} &=&
\dfrac{1}{\alpha^2}\left[-\delta^i_q\,\dfrac{\p \cH}{\p D_j} {}+{}
\left(\delta^{ij}\,\delta^s_p {}-{} \delta^i_p\,\delta^{js} \right)
\left(D^p\,\dfrac{\p^2 \cH}{\p D^s\, \p D^q} {}+{}
B^p\,\dfrac{\p^2 \cH}{\p B^s\, \p D^q} \right)
\right]
\;,
\\[2ex]
\dfrac{\p T^{ij}}{\p B^q} &=&
\dfrac{1}{\alpha^2}\left[-\delta^i_q\,\dfrac{\p \cH}{\p B_j} {}+{}
\left(\delta^{ij}\,\delta^s_p {}-{} \delta^i_p\,\delta^{js} \right)
\left(D^p\,\dfrac{\p^2 \cH}{\p D^s\, \p B^q} {}+{}
B^p\,\dfrac{\p^2 \cH}{\p B^s\, \p B^q} \right)
\right]
\;.
\end{array}
\end{eqnarray}
We have also
\begin{equation}
\begin{array}{rcl}
\dfrac{\p^2 \cH}{\p D^i \p D^j} &=&
\dfrac{\alpha^2}{\cH} \left[\delta_{ij} \left( 1 {}+{} \alpha^2\,\B^2 \right)
{}-{} \alpha^2\left(B_i\, B_j {}+{} E_i\,E_j\right)\right]
\quad,
\\[2ex]
\dfrac{\p^2 \cH}{\p B^i \p B^j} &=&
\dfrac{\alpha^2}{\cH} \left[\delta_{ij} \left( 1 {}+{} \alpha^2\,\D^2 \right)
{}-{} \alpha^2\left(D_i\, D_j {}+{} H_i\,H_j\right)\right]
\quad,
\\[2ex]
\dfrac{\p^2 \cH}{\p D^i \p B^j} &=&
\dfrac{\alpha^4}{\cH} \left[
2\,D_i\,B_j {}-{} \delta_{ij} \left( \D {}\cdot{} \B \right)
{}-{} B_i\, D_j {}-{} E_i\,H_j\right]
\quad.
\end{array}
\end{equation}
Using these formulas and formulas (\ref{Expr:EHDB}) we obtain from
(\ref{Exp:dTijdDB}) that
\begin{eqnarray}
\lb{Exp:dTijdDBm}
\begin{array}{rcl}
\dfrac{\p \st{n}{T}^{ij}}{\p D^q} &=&
-\delta^i_q\,\st{n}{E}^j {}+{} \delta^{ij}\,\st{n}{E}_q {}-{} \delta^j_q\,\st{n}{E}^i {}+{}
O\Bigl(1\Bigr/\rn{n}^4\Bigr)
\quad,
\\[2ex]
\dfrac{\p \st{n}{T}^{ij}}{\p B^q} &=&
-\delta^i_q\,\st{n}{H}^j {}+{} \delta^{ij}\,\st{n}{H}_q {}-{} \delta^j_q\,\st{n}{H}^i {}+{}
O\Bigl(1\Bigr/\rn{n}^4\Bigr)
\quad.
\end{array}
\end{eqnarray}
We substitute (\ref{Def:Pn}), (\ref{Expand:TijFirst}),
(\ref{Exp:dTijdDBm}) into relation (\ref{Cons:EM}\aref{b})
and make the integration for its right part.
Because the symmetry properties of element of integration,
the last integral is zero.
As result we obtain the following
generalized Lorentz equation for the trajectory of $n$-th dyon:
\begin{eqnarray}
\lb{Eq:trajectfirst}
\bar{\cE}\,\frac{\d }{\d x^0}\,\frac{\bfV{n}}{\sqrt{1 {}-{} \bfV{n}^2}}
&=&
\dn{n} \left(\tbD{n} {}+{} \bfV{n} {}\times{} \tbB{n}
\right) {}+{}
\bn{n} \left( \tbB{n} {}-{} \bfV{n} {}\times{} \tbD{n}
\right)
\quad.
\end{eqnarray}
As we see, the energy-momentum method for obtaining the trajectories
gives the value of mass which is equal to rest energy of the dyon.
This result fully conform with Einstein's principle for
equivalence of mass to energy.

In the article \cite{I1998HPA} we considered the potentials $A_\mu$
as unknown functions and used the boundary condition of type
(\ref{Cond:BoundAngle}) for obtaining the trajectory equation.
As result we had obtained another value of mass in initial
approximation ($\frac{3}{2}\,\bar{\cE}$)
than in the present article. Generally speaking, we have certain arbitrariness
for taking an initial trajectory of the singularity. Because a true trajectory
of the singularity is defined
by an exact solution only, we can obtain corrections to the initial
trajectory (in first and highest iteration steps) such that it will
appear as renormalization of mass. But  the method of the
present article is more suitable, because it gives the satisfaction
as energy-momentum conservation law (\ref{Cons:EM}) as
boundary condition (\ref{Cond:BoundAngle}) for any iteration step.

\section{Bidyon or an electromagnetic model of particle with spin}
\lb{Sec:Bidyon}

Let us consider the interaction of two dyon singularities
with identical electric and opposite magnetic charges:
$\dn{1} {}={} \dn{2}$, $\bn{1} {}={} -\bn{2}$. At first we use a cylindrical
coordinate system $\{z,\rho,\varphi \}$ such that dyons are on the axis $z$.
We believe also that the dyons velocities are directed with the axis $z$ and
the problem is symmetrical about the plane $z {}={} 0$.
This configuration is shown in figure \ref{Fig:twodyons}, where
$d {}={} \pm\bar{d}$, $b {}={} \pm\bar{b}$.
\FIGURE{
\lb{Fig:twodyons}
\begin{picture}(330,90)
\put(-65,-10){
\begin{picture}(330,100)
\put(60,30){\vector(1,0){330}}
\put(392,30){$z$}
\put(220,28){\line(0,1){4}}
\put(215.5,16){$0$}
\put(140,30){\circle*{3}}
\put(130,16){$-a$}
\put(140,40){\vector(-1,0){30}}
\put(100,44){$\bfV{1} {}={} {-\bfV{}}$}
\put(55,64){$\dn{1} {}={} d$}
\put(55,44){$\bn{1} {}={} b$}
\put(300,30){\circle*{3}}
\put(295.5,16){$a$}
\put(300,40){\vector(1,0){30}}
\put(295,44){$\bfV{2} {}={} \bfV{}$}
\put(345,64){$\dn{2} {}={} d$}
\put(345,44){$\bn{2} {}={} {-b}$}
\put(215.5,70){
\begin{picture}(50,50)
\put(0,0){\vector(0,1){20}}
\put(0,22){${\bf e}_\rho$}
\put(0,0){\vector(1,0){20}}
\put(22,0){${\bf e}_z$}
\put(0,0){\vector(-1,-1){10}}
\put(-25,-15){${\bf e}_\varphi$}
\end{picture}
}
\end{picture}
}
\end{picture}
\caption{Disposition of the two dyons in the
cylindrical coordinate system.}
}

We name this disposition of the singularities with appropriate field
configuration as bidyon.

According to our iterative method (see section \ref{Sec:PertMethod})
we take the sum of the fields for moving dyons (\ref{Sol:Dyonm}) with unknown
their trajectory $a(x^0)$ as initial approximation. That is we have
\begin{eqnarray}
\lb{Sol:InitBidyon}
\clD{}^{(0)} &=& \clD{1} {}+{} \clD{2}
\quad.
\end{eqnarray}
In the cylindrical coordinate system we have the following expressions
for the fields $\st{1}{\D},\,\st{2}{\D}$ and $\st{1}{\B},\,\st{2}{\B}$:
\begin{eqnarray}
\lb{BidyonFieldInitDB}
\begin{array}{lcl}
\left\{
\begin{array}{rcrcl}
\dfrac{\st{1}{D}_z}{d} &=& \dfrac{\st{1}{B}_z}{b} &=&
\dfrac{1}{\sqrt{1 {}-{} V^2}}\,\dfrac{z {}+{} a}{\rn{1}^3}
\\[2ex]
\dfrac{\st{1}{D}_\rho}{d} &=&
\dfrac{\st{1}{B}_\rho}{b} &=& \dfrac{1}{\sqrt{1 {}-{} V^2}}\,
\dfrac{\rho}{\rn{1}^3}
\\[2ex]
\dfrac{\st{1}{D}_\varphi}{b} &=&
\dfrac{\st{1}{B}_\varphi}{-d} &=& \dfrac{V}{\sqrt{1 {}-{} V^2}}\,
\dfrac{\rho}{\rn{1}\,\sqrt{\br^4 {}+{} \rn{1}^4}}
\end{array}
\right.
,\;
\left\{
\begin{array}{rcrcl}
\dfrac{\st{2}{D}_z}{d} &=& \dfrac{\st{2}{B}_z}{-b} &=&
\dfrac{1}{\sqrt{1 {}-{} V^2}}\,\dfrac{z {}-{} a}{\rn{2}^3}
\\[2ex]
\dfrac{\st{2}{D}_\rho}{d} &=&
\dfrac{\st{2}{B}_\rho}{-b} &=& \dfrac{1}{\sqrt{1 {}-{} V^2}}\,
\dfrac{\rho}{\rn{2}^3}
\\[2ex]
\dfrac{\st{2}{D}_\varphi}{b} &=&
\dfrac{\st{2}{B}_\varphi}{d} &=& \dfrac{V}{\sqrt{1 {}-{} V^2}}\,
\dfrac{\rho}{\rn{2}\,\sqrt{\br^4 {}+{} \rn{2}^4}}
\end{array}
\right.
,
\;\;\;\;\;
\end{array}
\end{eqnarray}
\bt{ll}
where&\quad $V {}\equiv{} \dfrac{\d a}{\d x^0}$ ,\quad
$\rn{1} {}={} \sqrt{\left(z^\prime {}+{} a^\prime\right)^2 {}+{} \rho^2 }$
,\quad
$\rn{2} {}={} \sqrt{\left(z^\prime {}-{} a^\prime\right)^2{}+{} \rho^2 }$
,\\[2ex]
&\quad
$z^\prime {}\equiv{} \dfrac{z}{\sqrt{1 {}-{} V^2}}
\quad,\qquad
a^\prime {}\equiv{} \dfrac{a}{\sqrt{1 {}-{} V^2}}$
\quad.
\et

\vskip1ex
The electromagnetic potentials don't use in our iterative method.
But from the initial approximation we can make some
suppositions about true field configuration of the potentials
for the appropriate exact solution. It is reasonable to take
that the both dyons into (\ref{Sol:InitBidyon}) have
the singular lines coinciding with the axis ${z}$.
That is we have
\begin{eqnarray}
\lb{BidyonFieldInitA}
\begin{array}{lcl}
\left\{
\begin{array}{rcrcl}
\dfrac{\st{1}{A}_z}{d} &=& \dfrac{\st{1}{\oA}_z}{-b} &=&
\dfrac{V}{\sqrt{1 {}-{} V^2}}\,\phi_0(\rn{1})
\\[2ex]
\dfrac{\st{1}{A}_\varphi}{b} &=&
\dfrac{\st{1}{\oA}_\varphi}{d} &=&
\phi_\varphi(z^\prime {}+{} a^\prime,\,\rho,\,\varphi)
\end{array}
\right.
,\qquad
\left\{
\begin{array}{rcrcl}
\dfrac{\st{2}{A}_z}{-d} &=& \dfrac{\st{2}{\oA}_z}{-b} &=&
\dfrac{V}{\sqrt{1 {}-{} V^2}}\,\phi_0(\rn{2})
\\[2ex]
\dfrac{\st{2}{A}_\varphi}{-b} &=&
\dfrac{\st{2}{\oA}_\varphi}{d} &=&
\phi_\varphi(z^\prime {}-{} a^\prime,\,\rho,\,\varphi)
\end{array}
\right.
.\qquad
\end{array}
\end{eqnarray}

Here we may take the infinite singular lines
with function $\phi_\varphi$ of type (\ref{Sol:TwoVarPhy}\aref{a}).
In this case the singular lines for the potential $\bA$ particularly
 cancel each other outside of the interval
between the dyons. This cancelling is complete as
$z {}\to{} \pm\infty$ or $a {}\to{} 0$.
If we also believe that at the singular points the field $\clD{}$ of
the bidyon solution are
equal to the field for the dyon solution $\clD{1}$ or $\clD{2}$,
we must believe that the exact solution has
the potential $\bA$ with the function $\phi_\varphi$
of type (\ref{Sol:TwoVarPhy}\aref{b}) at the singular
points.
Thus we can presume
that the potential $\bA$ for the associated exact solution has
the singular line segment between the dyons only.
That is we have a singular string segment
(for $\bA$) with the dyons at the ends.

Let us calculate the full
 angular momentum for field configuration (\ref{Sol:InitBidyon}). We use
definition (\ref{Def:VectAngMom}) for full three-dimensional space.
Because of the symmetry property of elements of integration into
(\ref{Def:VectAngMom}), for this case we have
$M_\rho {}={} M_\varphi {}={} 0$,
\begin{eqnarray}
\lb{Def:Mz}
M_z &=& -\frac{1}{4\pi}
\int \cP_\varphi\,\rho \;(\rho\, \d z \d\rho\d\varphi)
\quad.
\end{eqnarray}
We can easy obtain the following expression:
\begin{eqnarray}
\lb{Expr:Pphi}
\cP_\varphi &=& \frac{-4\,a\,b\,d\,\rho}%
{\rn{1}^3\,\rn{2}^3\,\left(1 {}-{} V^2\right)}
\quad.
\end{eqnarray}

Now we introduce new variables of integration $\{\xi,\zeta,\varphi\}$
that appropriate to the bispherical coordinate system:
\begin{eqnarray}
\lb{Trans:bispher}
z^\prime {}={} \frac{a^\prime\,\sinh{\xi}}{\cosh{\xi}{}-{} \cos{\zeta}}
\quad,\qquad
\rho {}={} \frac{a^\prime\,\sin{\zeta}}{\cosh{\xi}{}-{} \cos{\zeta}}
\quad.
\end{eqnarray}
We have the bispherical element of value
\begin{eqnarray}
\lb{Expr:Value}
(\rho\, \d z^\prime \d\rho\d\varphi) &=&
\dfrac{a^{\prime 3}\,\sin{\zeta}\;\d\xi\,\d\zeta\,\d\varphi}%
{\left(\cosh{\xi}{}-{} \cos{\zeta}\right)^3}
\quad.
\end{eqnarray}
We have also the following expression for $\rn{1}$, $\rn{2}$:
\begin{eqnarray}
\lb{Expr:rr}
\rn{1} {}={} \frac{\sqrt{2}\,a^\prime\,\exp{\left(\xi/2\right)}}%
{\sqrt{\cosh{\xi}{}-{} \cos{\zeta}}}
\quad,\qquad
\rn{2} {}={} \frac{\sqrt{2}\,a^\prime\,\exp{\left(-\xi/2\right)}}%
{\sqrt{\cosh{\xi}{}-{} \cos{\zeta}}}
\quad.
\end{eqnarray}

Substituting (\ref{Trans:bispher}), (\ref{Expr:rr}) into (\ref{Expr:Pphi})
and introducing the variable $\uz {}={} \cos{\zeta}$,
we obtain the following expression:
\begin{eqnarray}
\lb{Expr:Pphib}
\cP_\varphi &=&
-b\,d\,\dfrac{\left(1 {}-{} V^2\right)^{\frac{3}{2}}}{2\,a^4}
\left( \cosh\xi {}-{} \uz \vphantom{\bV2{}}\right)^2 \sqrt{1 - \uz^2}
\quad.
\end{eqnarray}
And substituting (\ref{Trans:bispher}), (\ref{Expr:Value}), (\ref{Expr:Pphib})
into (\ref{Def:Mz})
we obtain the following expression for
$z$-component of the full angular momentum:
\begin{eqnarray}
M_z &=&
\frac{b\,d}{8\pi}\,
\int\limits^\pi_{-\pi} \d\varphi
\int\limits^\infty_{-\infty} \d\xi
\int\limits^1_{-1}
\frac{\left(1 {}-{} \uz^2\right)}{\left(\cosh{\xi} {}-{} \uz\right)^2}
\;\d\uz
\quad.
\end{eqnarray}
Making firstly the integration over $\uz$ and $\xi$ in the finite
limits $[-\bar{\xi},\,\bar{\xi}]$ we obtain
\begin{equation}
\int\limits^\infty_{-\infty} \d\xi
\int\limits^1_{-1}
\frac{\left(1 {}-{} \uz^2\right)}{\left(\cosh{\xi} {}-{} \uz\right)^2}
\;\d \uz
{\,}={\,} \lim_{\bar{\xi} {}\to{} \infty}
\left[ 4\left(\ln
\frac{\cosh{\bar{\xi}} {}+{} 1}{\cosh{\bar{\xi}} {}-{} 1}
\right) \sinh{\bar{\xi}}\,
\right]
{\,}={\,} 8
\quad.
\end{equation}
Thus we have
\begin{eqnarray}
\lb{Expr:Angmom}
M_z &=& 2\,b\,d
\quad.
\end{eqnarray}
As we see, the full angular momentum is independent of the
interdyonic distance and their speed!
It is evident that this result is typical for any
electrodynamics model, i.e. the form of the
function $\cH (\D,\B)$ is immaterial.

The property, that the full angular momentum is independent of
the internal movement parameters,
permits a consideration of the bidyon
as an electromagnetic model for charged particle with spin.
The charge of this particle is $2\,d$ and its spin is
equal to $|M_z|$. We set $2\,\bd {}={} e$, were $e$ is the absolute
value of the electron charge, and $|M_z| {}={} \dfrac{\hbar}{2}$.
Thus we have
\begin{eqnarray}
\bar{b}\,e {}={} \frac{\hbar}{2}
\quad\Longrightarrow\quad
\bar{b} {}={} \frac{e}{2}\,\frac{\hbar}{e^2} {}={}
\frac{e}{2}\,\frac{1}{\bar{\alpha}}
\quad\Longrightarrow\quad
\frac{\,\bd\,}{\,\bar{b}\,} {}={} \bar{\alpha}
\quad,
\lb{Def:baralpha}
\end{eqnarray}
where $\bar{\alpha} {}={} e^2/\hbar {}\approx{} 0.00729735$
is the fine structure constant.

Now we shall find the trajectory $a(x^0)$ in initial approximation.
For this purpose we use the integral energy conservation law
for full three-dimensional space. The full energy
of initial field configuration (\ref{Sol:InitBidyon})
is depend on the distance between the dyons $a$ and
on the speed $V$: ${\cE} {}={} {\cE}(a,\,V)$.
On the trajectory $a(x^0)$ the full energy is constant.
Thus we can obtain the trajectory with help the following formula:
\begin{eqnarray}
\lb{Sol:Motion}
x^0 &=&
\dint{a}{a(0)}\dfrac{\d \tilde{a}}{V(\tilde{a},\,{\cE})}
\quad,
\end{eqnarray}
where $V(a,\,{\cE})$ is the inverse function to the function
${\cE}(a,\,V)$.

It is evident that we have a family of the trajectories for all possible
values of the energy.

But in fact, the energy depend on square of the speed. Thus
we can obtain the acceleration from the function ${\cE}(a,V^2)$
by the following way:
\begin{equation}
\lb{Expr:Acc}
\dfrac{\d{\cE}}{\d x^0} {}={} \dfrac{\p \cE}{\p a}\,V {}+{}
\dfrac{\p \cE}{\p V^2}\,2\,\dfrac{\d V}{\d x^0}\,V {}={} 0
\qquad \Longrightarrow\qquad
\dfrac{\d V}{\d x^0} {}={} -\dfrac{1}{2}\,\dfrac{\p \cE}{\p a}\Biggl/
\dfrac{\p \cE}{\p V^2}
\quad.
\end{equation}

Let us obtain the function ${\cE}(a,\,V^2)$. We shall use a numerical
calculation for the energy integral. But at first we obtain some interesting
approximate formula for the energy in the case of very small distance
between the dyons and their speed is not very close to the speed of light.
We substitute the field $\clD{}^{(0)}$
into expression for $\cH$ (2.7) and use formulas
 for bispherical coordinates (9.6) and (9.8).
We obtain that as $a^\prime {}\to{} 0$ the main term of the radicand is
$\alpha^4 \left(\cP_\varphi\right)^2$, which is proportional to
$\left(a^\prime\right)^{-8}$.
Thus for very small values of $a^\prime$ we have the following formula:
\begin{eqnarray}
{\cE} &=& \frac{1}{4\pi}
\int \left|\cP_\varphi \right|\;(\rho\, \d z \d\rho\d\varphi) {\,}={\,}
\frac{\bar{b}\,\bd}{8\pi\,a^\prime}\,
\int\limits^\pi_{-\pi} \d\varphi
\int\limits^\infty_{-\infty} \d\xi
\int\limits^1_{-1}
\frac{\sqrt{1 {}-{} \uz^2}}{\left(\cosh{\xi} {}-{} \uz\right)}
\;\d\uz
\quad.
\end{eqnarray}
Making firstly the integration over $\xi$ we obtain\qquad
${\displaystyle
\int\limits^\infty_{-\infty} \d\xi
\int\limits^1_{-1}
\frac{\sqrt{1 {}-{} \uz^2}}{\left(\cosh{\xi} {}-{} \uz\right)}
\;\d \uz
{\,}={\,}}$
\begin{equation}
{\,}={\,}
2\int\limits^1_{-1}
\left[
\arctan{\left(\sqrt{\frac{1 {}+{} \uz}{1 {}-{} \uz}}\,\right)}
{}-{}
\arctan{\left(-\sqrt{\frac{1 {}+{} \uz}{1 {}-{} \uz}}\,\right)}
\right] \d\uz
{\,}={\,}
2\,\pi\quad.
\end{equation}
As result we have the following formula:
\begin{eqnarray}
\lb{Expr:EnergySmalla}
{\cE} &=& \frac{\pi\,\bar{b}\,\bd\,\sqrt{1 {}-{} V^2}}{2\,a}
\quad.
\end{eqnarray}
As we see, in the case of $V {}\neq{} 1$ this formula gives
that ${\cE} {}\to{} \infty$ as $a {}\to{} 0$! Moreover
the effective mass of this system is negative, in the sense
that the acceleration (see (\ref{Expr:Acc}))
is directed to the region with a greater rest energy!

Notice, the sum $\clD{1} {}+{} \clD{2}$
are considered here as initial approximation, in particular,
for very small distance between the singularities
($2\,a {}\ll{} \br$).
However, as indicated above (see the end of section \ref{Sec:PertMethod})
this may be rightful, because absolute value of the dyon's fields $\D,\,\B$
are infinity at the singular point.
Thus we can hope that our iterative procedures (\ref{iter1}) or
(\ref{iter2}) with this initial approximation can give
an approximated solution. In any case the investigation of
field configuration (\ref{Sol:InitBidyon}) may give leading
considerations for obtaining the solution.

Now we present the results of numerical
calculations\footnote{Here we had used {\tt FORTRAN} program with {\tt NAG}
library (function {\tt D01DAF} for integration in two dimensions).}
for the function ${\cE}(a,\,V^2)$.
We used the dimensional system
in which $\alpha {}={} 1$ and $\br {}={} 1$. In this
dimensional system the energy of $\clD{}^{(0)}$
for $a {}={} \infty$ and $V {}={} 0$
is $2\,\bcE {}={} 4\beta {}/{} 3 {}\approx{} 2.4721$
(see (\ref{Def:bcE})), and equality $a {}={} 1$
implies that $a {}={} \br$ in other dimensional system. According to
(\ref{Def:baralpha}), (\ref{Def:br}) we have $d {}={} \bar{\alpha}\,b$,
$b {}={} \left(1 {}+{} \bar{\alpha}^2\right)^{-\frac{1}{2}}$
in this dimensional system.
We made the calculation for $a {}={} 0.00004\div 4$ and
$V^2 {}={} 0\div 0.95$.
In order to make the best convergence
of the integral (both near the singular points and as
$\rn{1},\,\rn{2} {}\to{} \infty$), we divide  the space of integration
on two part. In the first part, which is bounded by the sphere with radius
$20\div 40$, we used the bispherical coordinates. And outside this sphere
but inside the sphere with radius $200\div 1000$ we used the spherical
coordinates.
The precision for obtained results of calculations is
no more than $5\%$.

As noted above the effective mass for the internal movement of
the dyons in $\clD{}^{(0)}$
may be negative for very small distance between the dyons. But, of course,
as $a {}\to{} \infty$ this mass is positive. Thus there are an
inflectional line in coordinates $(a,\,V^2)$ on which the mass is equal to zero.
We denote the point of intersection this line with the axes $a$ as $\ua$.
That is
\begin{eqnarray}
\left.\dfrac{\p {\cE}(a,V^2)}{\p V^2}\right|_{%
{ \begin{array}{rcl} a &=&\ua\\[-0.15em]
V &=&  0\end{array}}}
&=& 0
\quad.
\end{eqnarray}
According our numerical calculations $\ua {}\approx{} 4.6\cdot 10^{-3}$
and value of the energy ${\cE} (\ua,0)$ is near $4\beta {}/{} 3$.
It is possible that this is her exact value. This point $(\ua,0)$
in some sense is dual to the point $(\infty,0)$ where the energy
is $\cE {}={} 2\,\bcE$.
We shall also see this duality
below in the context of obtained trajectories $a(x^0)$.

The numerical calculations give that for $a {}<{} \ua$ and $V {}={} 0$
the values of the energy practically equal to ones given
by formula (\ref{Expr:EnergySmalla}).
But this agreement disappears as the
speed $V$ increase. The appropriate results show in figure
\ref{Fig:energy0}.
(The values calculated by formula (\ref{Expr:EnergySmalla})
show with discontinuous line.)
\FIGURE{
\begin{picture}(330,95)
\put(-65,-175){
\epsfig{file=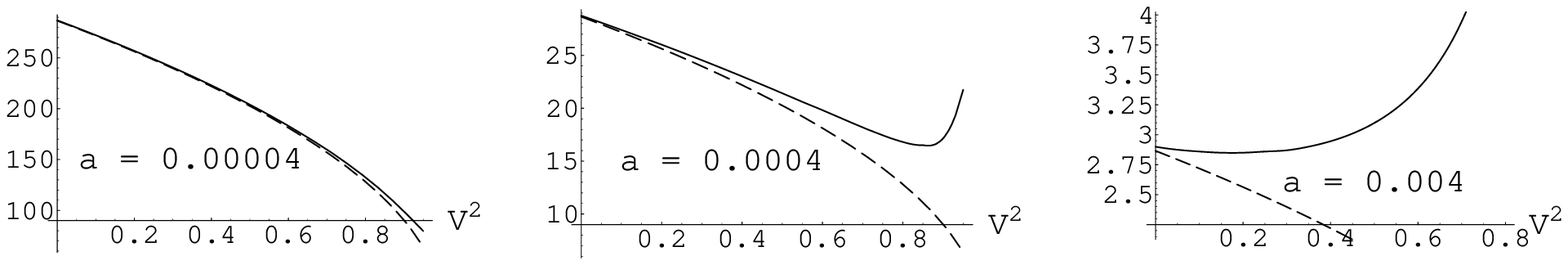,width=90ex}%
}
\put(-40,82){$\cE$}
\put(107,82){$\cE$}
\put(268,82){$\cE$}
\end{picture}
\caption{Energy of ${\cal D}^{(0)}$ for $a {}<{} \ua$
(approximation by formula (\ref{Expr:EnergySmalla})).}
\label{Fig:energy0}
}

The best agreement with the numerical results is given by addition the
term $C\,\frac{V^2}{1 {}-{} V^2}$ to expression (\ref{Expr:EnergySmalla})
for the energy. Here the coefficient $C$ is chosen
from the condition
${\cal E}(\ua,0) {}={} 2\,\bcE$.
Thus we have the following approximate formula for the energy:
\begin{eqnarray}
\lb{Expr:EnergySmallaM}
{\cE} &=& \frac{\pi\,\bar{b}\,\bd\,\sqrt{1 {}-{} V^2}}{2\,a} {}+{}
\dfrac{2}{3}\,\beta\,\alpha^{-2}\,\br^3\,\dfrac{V^2}{1 {}-{} V^2}
\quad.
\end{eqnarray}
According this formula we have the following expression:
\begin{eqnarray}
\lb{Expr:ua}
\ua &=& \frac{3\,\alpha^2\,\pi\,\bar{b}\,\bd}{8\,\beta\,\br^3}
\quad.
\end{eqnarray}
In our dimensional system this formula gives $\ua {}={} 4.63673\cdot 10^{-3}$
that good agree with the numerical result.
By addition the terms
$\sim\!\!\left(\frac{V^2}{1 {}-{} V^2}\right)^n$,
where
$n {}={} 2,3,4$, to formula (\ref{Expr:EnergySmallaM})
we can obtain a formula,
which gives very good approximation for the results of numerical calculations
(for $a {}<{} \ua$ and $V^2 {}<{} 0.95$).
This formula has the following form:
\begin{equation}
\lb{Expr:EnergySmallaGood}
\;\;{\cE} {}={}
{\frac{0.0115\,{\sqrt{1 - {V^2}}}}{a}} +
  {\frac{1.236\,{V^2}}{1 - {V^2}}} -
  {\frac{0.083\,V^4}{{{\left( 1 - {V^2} \right) }^2}}} +
  {\frac{0.007\,{V^6}}{{{\left( 1 - {V^2} \right) }^3}}} -
  {\frac{0.0002\,V^8}{\left( 1 - {V^2} \right)^4}}
\;\;.\quad
\end{equation}
Here we change the two factors in formula
(\ref{Expr:EnergySmallaM}) for their numerical values
in our dimensional system. In figure \ref{Fig:energya}
is shown this good agreement between the results of numerical
calculations and values calculated by formula
(\ref{Expr:EnergySmallaGood}).
\FIGURE{
\begin{picture}(330,95)
\put(-65,-175){
\epsfig{file=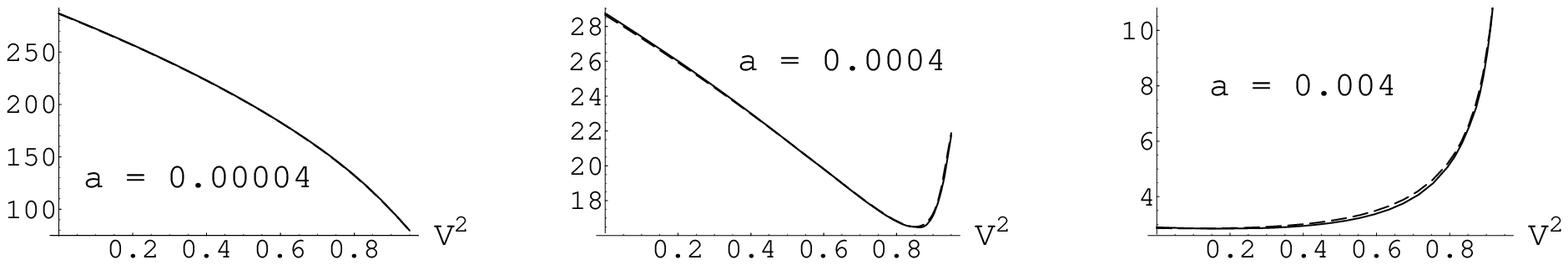,width=90ex}%
}
\put(-37,82){$\cE$}
\put(113,82){$\cE$}
\put(266,82){$\cE$}
\end{picture}
\caption{Energy of ${\cal D}^{(0)}$ for $a {}<{} \ua$
(approximation by formula (\ref{Expr:EnergySmallaGood})).}
\label{Fig:energya}
}\\
It is evident that formula
(\ref{Expr:EnergySmallaGood}) gives the same expression for $\ua$
(\ref{Expr:ua}).

In the region of $a {}>{} \ua$ the function ${\cE}(a,V^2)$
has a minimum. The appropriate plot for $V {}={} 0$ is shown in figure
\ref{Fig:energy1}.
\FIGURE{
\begin{picture}(330,140)
\put(-110,-280){
\epsfig{file=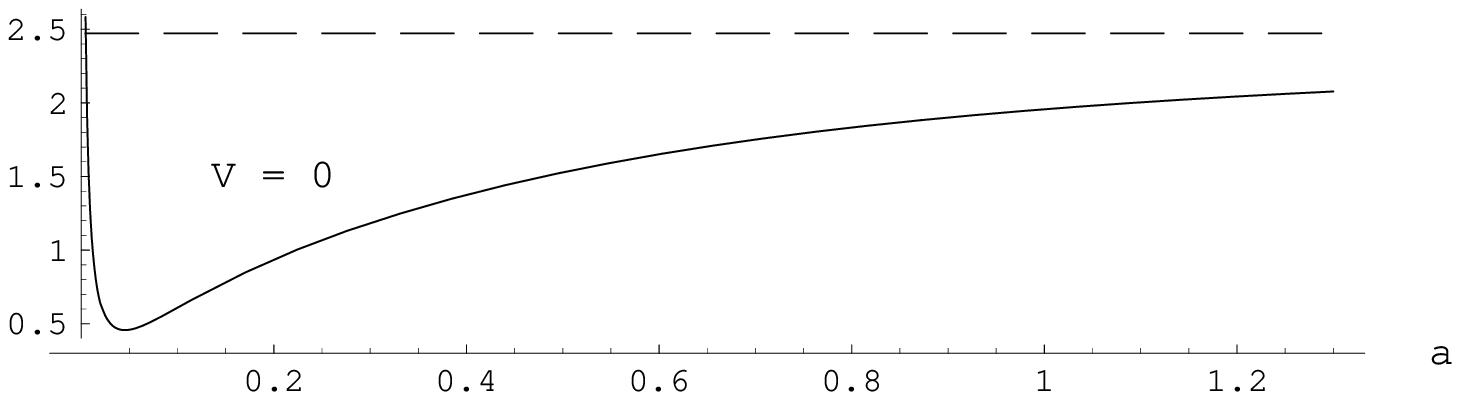}%
}
\put(-25,117){$\cE$}
\end{picture}
\caption{Energy of ${\cal D}^{(0)}$ for middle values of $a$ and $V {}={} 0$.}
\label{Fig:energy1}
}

As we see, the energy falls very quick and then it increases very slow
as the values of $a$ increase. Such dependence is convenient to show
by the plot with logarithmic scale on $a$. In figure \ref{Fig:energy2}
is shown the function ${\cE}(\log_{10} a,\,V^2)$ for
 $V^2 {}={} 0\div 0.6$ with step $0.05$.
\FIGURE{
\begin{picture}(330,195)
\put(-20,-400){\epsfig{file=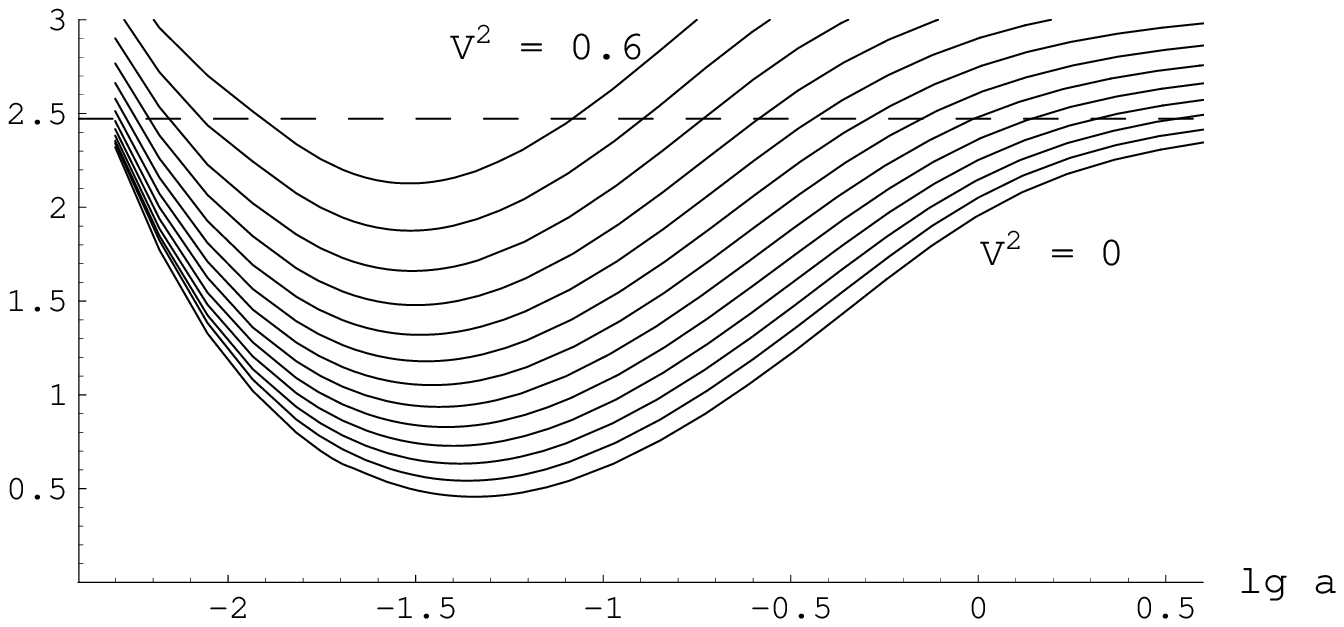}%
}
\put(2,185){$\cE$}
\end{picture}
\caption{The energy in logarithmic scale of $a$
and $V^2 {}={} 0\div 0.6$ with step $0.05$.}
\label{Fig:energy2}
}
Here we have the minimal value of the energy
${\cE}_m {}\approx{} 0.46$ for $a {}={} a_m {}\approx{} 0.045$.
For very small value of deviation $|a {}-{} a_m|$ we have the following
approximate formula:
\begin{eqnarray}
\lb{Exmr:cEm}
{\cE}&=& {\cE}_m {}+{} 128\left(a {}-{} a_m\right)^2 {}+{} 1.65\,V^2
\quad.
\end{eqnarray}

Now we investigate the trajectories for various values of the energy.
With the numerical calculations we have obtained the function
$V(a,\cE)$ for the region of $a$ which is shown in figure
\ref{Fig:energy2}. The appropriate results are shown in figure
\ref{Fig:energy3}.
\FIGURE{
\begin{picture}(330,200)
\put(-20,-400){\epsfig{file=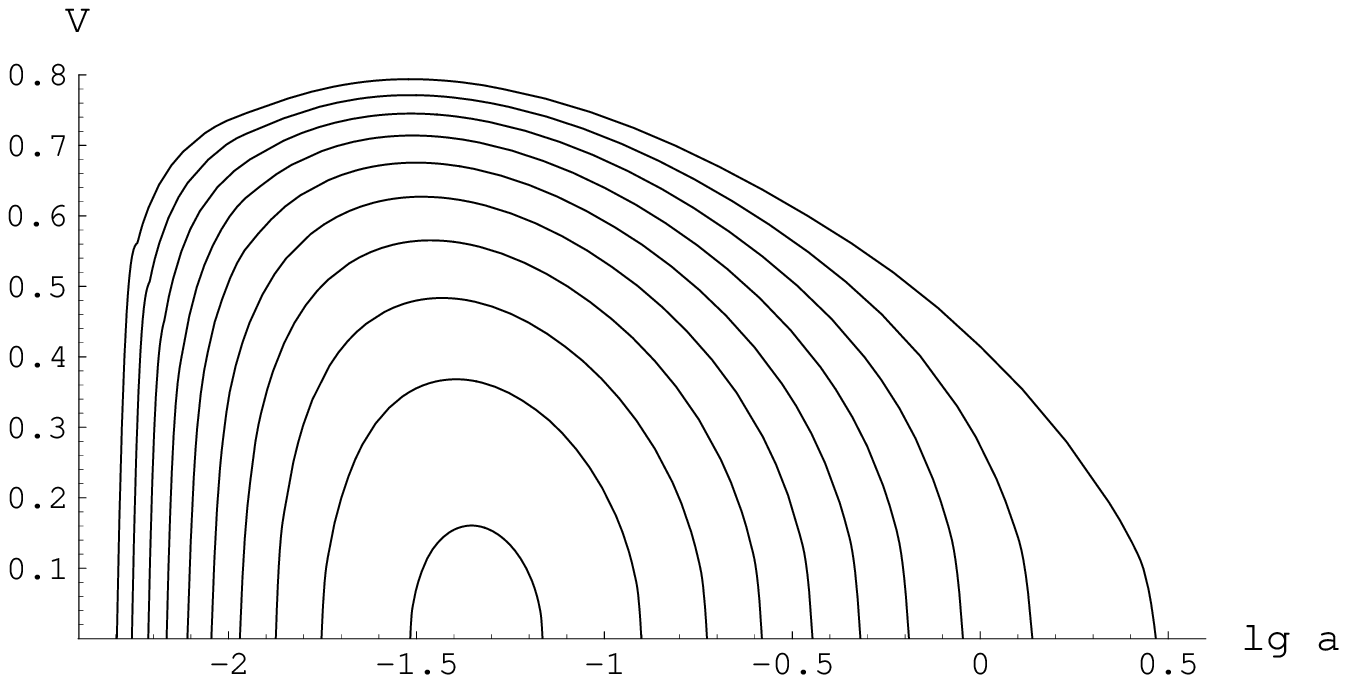}%
}
\end{picture}
\caption{The function $V(\log_{10} a,\,{\cE})$ for
${\cE} {}={} 0.5\div 2.3$ with step $0.2$.}
\label{Fig:energy3}
}
As we see, here we have a periodical motion. According to formula
(\ref{Sol:Motion}) the period of this motion is given by the formula
\begin{eqnarray}
\lb{Sol:Period}
T &=&
2\,\dint{a_{max}}{a_{min}}\dfrac{\d \tilde{a}}{V(\tilde{a},\,{\cE})}
\qquad \mbox{and} \qquad
\omega {\,}\equiv{\,} \frac{2\pi}{T}
\quad.
\end{eqnarray}
For the minimal value of the energy $\cE_m$ we calculate the circular
frequency $\omega$ using formula (\ref{Exmr:cEm}). The appropriate
value is $8.825$. And for
values of the energy from $0.5$ to $2.3$ with step $0.1$ we
numerically calculate the frequency with help of formula
(\ref{Sol:Period})\footnote{In fact we numerically calculated
the integral from $a_{min} {}+{} \delta a_1$ to
$a_{max} {}-{} \delta a_2$. For the regions
$[a_{min},a_{min} {}+{} \delta a_1]$, $[a_{max} {}-{} \delta a_2,a_{max}]$
we approximated the function $\cE(a, V^2)$ by a linear dependence on $a$,
$V^2$ and analytically calculate the appropriate integral.}. As result
we obtain the function $\omega (\cE)$. The obtained dependence
$\omega (\cE)$ is shown in figure \ref{Fig:energy4}.
The dual points $(a,V) {}={} (\ua,0)$ and $(\infty,0)$ are unstable
points of rest with the energy $2\,\bcE$. The appropriate
period is infinity and the frequency is zero.
\FIGURE{
\begin{picture}(330,180)
\put(10,-330){\epsfig{file=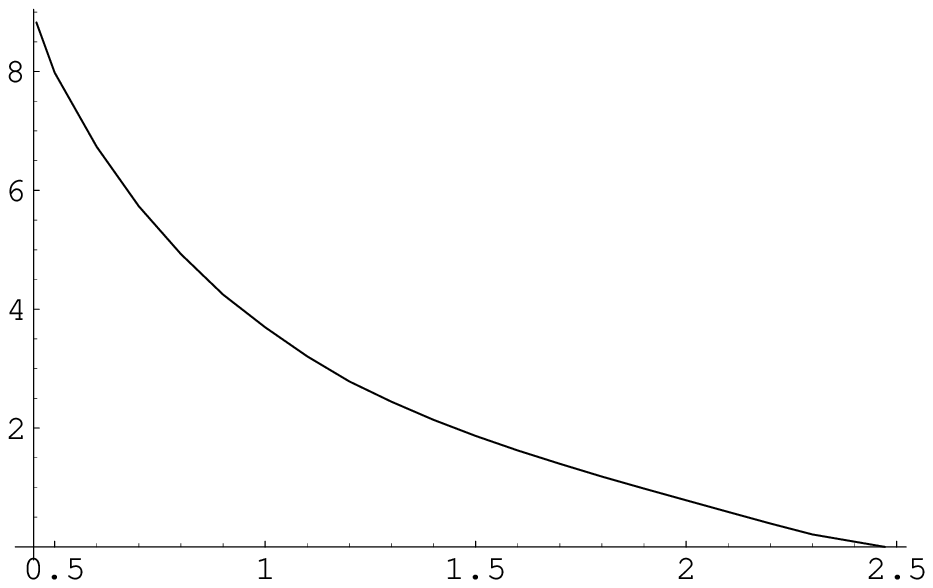}%
}
\put(15,182){$\omega$}
\put(285,15){$\cE$}
\end{picture}
\caption{The function $\omega ({\cE})$ of
the periodical motion for ${\cE} {}<{} 4\beta {}/{} 3$.}
\label{Fig:energy4}
}

Now we consider the movement in the region $a {}<{} \ua$.
The lines with constant energy for this region are shown in
figure \ref{Fig:energy5}.
\FIGURE{
\begin{picture}(330,210)
\put(-20,-390){\epsfig{file=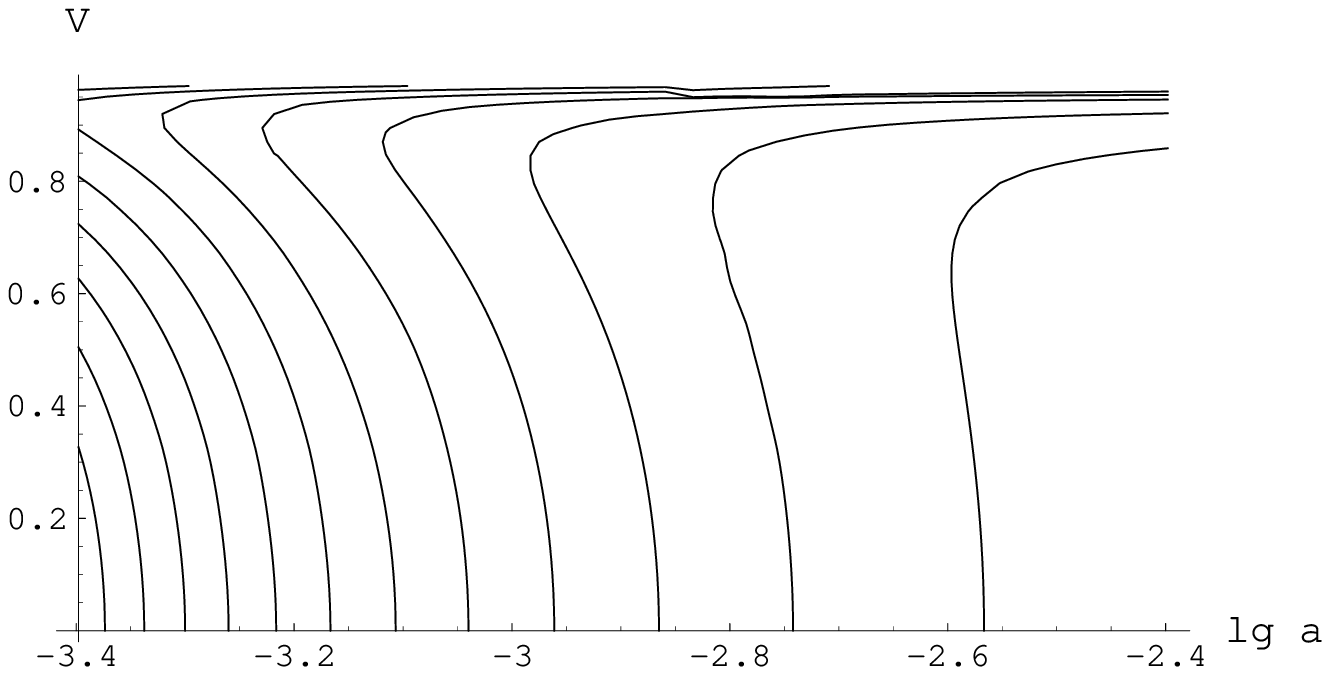}%
}
\end{picture}
\caption{Lines with constant energy for $a {}<{} \ua$.}
\label{Fig:energy5}
}
Let us consider the line in figure
\ref{Fig:energy6} that correspond to sole value of the energy.
\FIGURE{
\begin{picture}(330,230)
\put(-20,-420){\epsfig{file=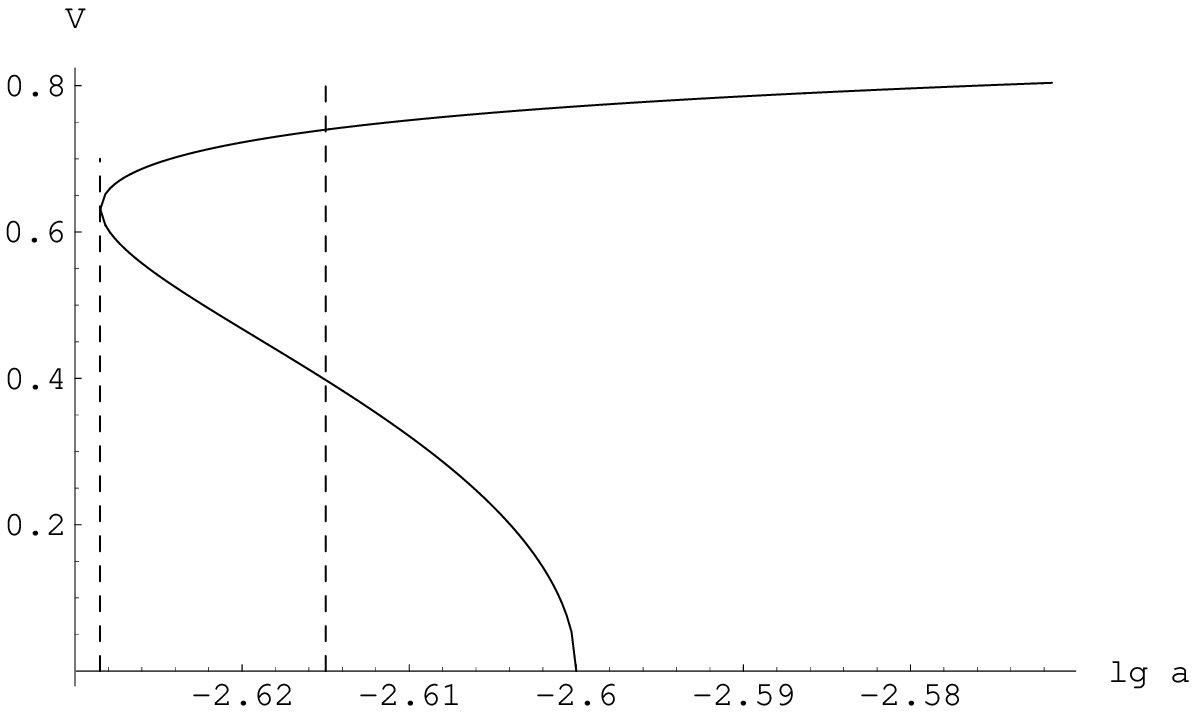}}
\put(4,0){$\ae$}
\put(141,0){$\ab$}
\end{picture}
\caption{Line with constant energy for $a {}<{} \ua$.}
\label{Fig:energy6}
}
As we see in this figure, if for an initial instant of time
the speed is zero ($V {}={} 0$),
then the dyons begin to approach each other
from the points $a {}={} \ab (\cE)$ and quick gather a high speed.
According to the all formulas for the energy in the region $a {}<{} \ua$
( (\ref{Expr:EnergySmalla}), (\ref{Expr:EnergySmallaM})
and (\ref{Expr:EnergySmallaGood}) )
 we have the following expression:
\begin{eqnarray}
\lb{Expr:ab}
\ab &=& \dfrac{\pi\,\bar{b}\,\bd}{2\,\cE}
\quad.
\end{eqnarray}

But at any intermediate point
between $\ae$ and $\ab$ in figure \ref{Fig:energy6} the dyons
may have two different values of the speed for one value of the energy.
In principle,  the system may jump from one branch of the
constant energy line to another. And in the initial approximation
there are not any criterion for choosing between these branches. But in first
approximation we take into account a radiation of the dyons.
Because an amplitude of the radiation depend on acceleration,
 we can propose that this radiation will prevent for the jumping
(with infinite acceleration). Thus we believe that the bidyon system
in initial approximation
moves from the point $a {}={} \ab$, $V {}={} 0$ to the point $\ae$
on the bottom (in the figure) branch of the constant energy line.
At the point $a {}={} \ae$
the acceleration on the trajectory is infinity and direction of the
velocity is changed to opposite one. From the point $\ae$
the dyons may move on the top branch ($a {}\to{} \infty$)
or come back on the bottom one.
Of course, near this point
the radiation plays a significance role for the movement.
Thus we can suppose that the bidyon dynamical system stop short of
the point with infinite acceleration and come back on
the branch with less speed, that is the bottom branch.
In this case we have the periodical motion between the points $\ab$
and $\ae$.

We shall numerically calculate the period for this periodical motion.
But at first we shall have obtained some analytical result for
an idealized case. If we have a dynamical system with energy given
by formula (\ref{Expr:EnergySmalla}), then
we have her trajectory with the formula
\begin{equation}
\lb{Sol:MotionA}
x^0 {}={} \dint{a}{\ab} \left[ 1 {}-{}
\left(\dfrac{2\,\tilde{a}\,\cE}{\pi\,\bar{b}\,\bd}\right)^2\right]^{-\frac{1}{2}}
\d \tilde{a}
{}={}
\left.\dfrac{\pi\,\bar{b}\,\bd}{2\,\cE}\,
\arcsin\left(\dfrac{2\,\tilde{a}\,\cE}{\pi\,\bar{b}\,\bd}\right)\right|^a_{\ab}
\quad.
\end{equation}
Using formula (\ref{Expr:ab}) for $\ab$ we can obtain
the following trajectory:
\begin{equation}
\lb{TrajectIdeal}
a {}={} \dfrac{2}{\omega}\left|\cos \dfrac{\omega}{2} x^0\right|
\quad,\qquad
\omega {}={} \dfrac{4\,\cE}{\pi\,\bar{b}\,\bd}
{}={} \dfrac{16\,\cE}{\pi\,\hbar}
\quad.
\end{equation}
As we see, in this case the energy is proportional to the frequency!
The system moves from the point $a {}={} \ab$  to the point $a {}={} 0$
and come back. In keeping with the conservation of full angular momentum,
the dyons may not change places. At the point $a {}={} 0$ the speed
is equal to the speed of light $V {}={} 1$, for this idealized system.

Now we present a dependence $\omega (\cE)$ that numerically calculated
by formula (\ref{Expr:EnergySmallaGood}) for the energy.
As noted, this formula very good approximate the numerically calculated
function $\cE(a,V)$.
In figure \ref{Fig:energy7} we have the plot for numerically calculated
function $\omega (\cE)$ (continuous line), asymptote
$\cE {}={} 2\,\bcE$, and dependence $\omega (\cE)$
for idealized motion (\ref{TrajectIdeal}) (discontinuous lines).
\FIGURE{
\begin{picture}(330,200)
\put(0,-375){\epsfig{file=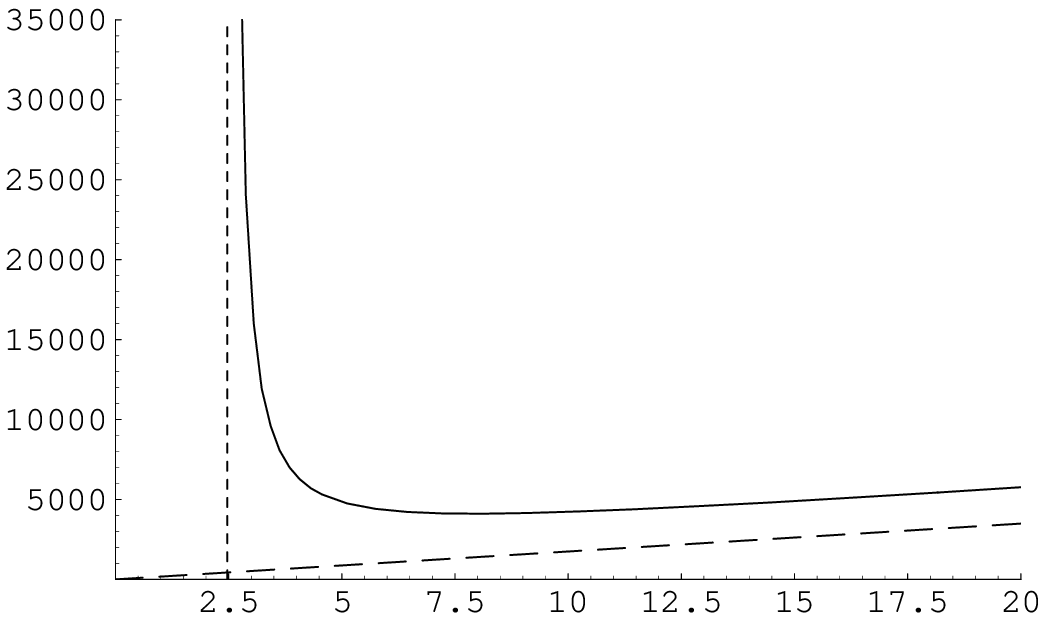}}
\put(28,190){$\omega$}
\put(310,10){$\cE$}
\end{picture}
\caption{Function $\omega (\cE)$ for the periodical motion in $a {}<{} \ua$.}
\label{Fig:energy7}
}\\
As we see, $\omega {}\to{} \infty$ ($T {}\to{} 0$) when the energy
approached $2\,\bcE$ from above, in contrast to the case when
the energy approached $2\,\bcE$ from below.

We can derive also a formula that will better approximate for the numerically
calculated function $\omega (\cE)$. Let us take  movement
(\ref{Sol:MotionA}) of idealized system (\ref{Expr:EnergySmalla}).
But we shall suppose that the system stop short of the point $a {}={} 0$
and come back from some point $\ae {}\neq{} 0$. We shall
have obtained this point from approximate formula (\ref{Expr:EnergySmallaM}).
This formula considered as equation with their associated equation
$\dfrac{\p\cE}{\p V^2} {}={} 0$ are the system for obtaining of the point
 $(\ae, \Ve^2)$ with infinity acceleration.\\[1ex]
The appropriate solution is
\begin{equation}
\ae {}={} \dfrac{3\sqrt{3}\,\pi\,\bar{b}\,\bd\,\sqrt{\bcE}}%
{4\,\left(\cE {}+{} \bcE\right)^{\frac{3}{2}}}
\quad,\qquad
\Ve^2 {}={} \dfrac{\cE {}-{} 2\,\bcE}{\cE {}+{} \bcE}
\quad.
\end{equation}
Using expression (\ref{Sol:MotionA}) we have the period for this
restricted idealized motion in the following form:
\begin{eqnarray}
T &=& \dfrac{\pi\,\bar{b}\,\bd}{\cE} \left[\dfrac{\pi}{2} {}-{}
\arcsin\left(\dfrac{2\,\ae\,\cE}{\pi\,\bar{b}\,\bd}\right) \right]
\quad.
\end{eqnarray}
And for the circular frequency we have
\begin{eqnarray}
\lb{Expr:omegaappr}
\omega &=& \dfrac{2\,\cE}{\bar{b}\,\bd}\left/
\left\{\dfrac{\pi}{2} {}-{} \arcsin\left[
\dfrac{3\,\sqrt{3\,\bcE}\,\cE}{2\left(
\cE {}+{} \bcE\right)^{\frac{3}{2}}}
\right]
\right\}
\right.
\quad.
\end{eqnarray}
In figure \ref{Fig:energy8} is shown this function
$\omega (\cE)$ (discontinuous curve) with the numerically calculated
function according to formula (\ref{Expr:EnergySmallaGood})
for the energy (continuous curve).
\FIGURE{
\begin{picture}(330,200)
\put(0,-375){\epsfig{file=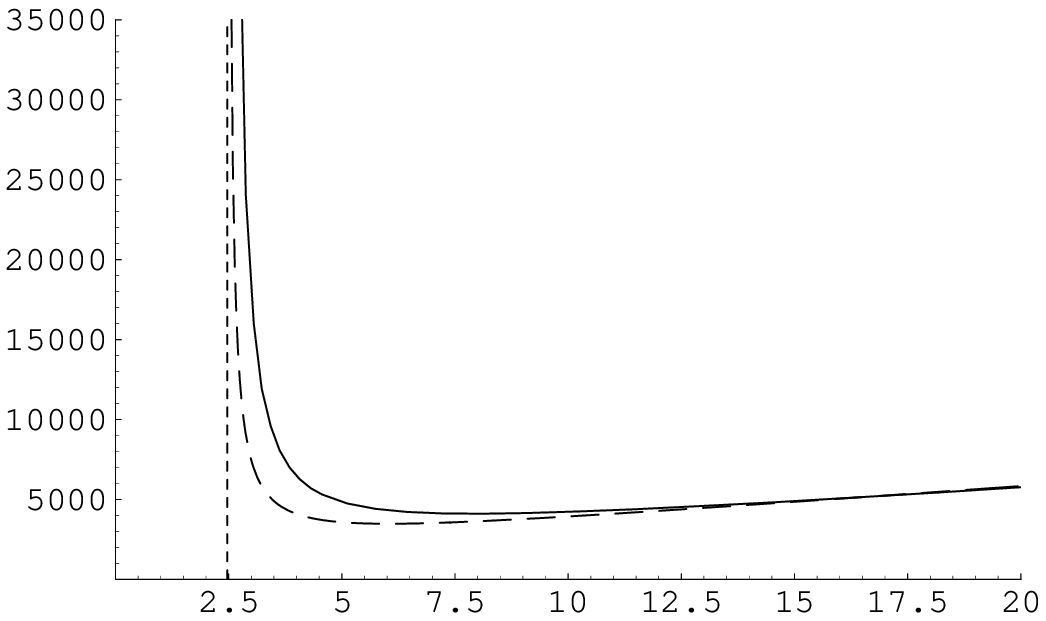}}
\put(28,190){$\omega$}
\put(310,10){$\cE$}
\end{picture}
\caption{The function $\omega (\cE)$ for $a {}<{} \ua$ with
its approximation.}
\label{Fig:energy8}
}\\
As we see, there are some distinction
between the plots near $\cE {}={} 2\,\bcE$. But for superior value
of the energy we have good agreement between the plots. For high
values of the energy formula (\ref{Expr:omegaappr}) gives
\begin{eqnarray}
\lb{Expr:omegaappr1}
\omega &\approx& \dfrac{4\,\cE}{\pi\,\bar{b}\,\bd} {}+{}
\dfrac{12\,\sqrt{3\,\bcE}}{\pi^2\,\bar{b}\,\bd}\,\sqrt{\cE}
\quad.
\end{eqnarray}

Thus we have considered the internal movement
of the bidyon for initial approximation, which is the sum of the
dyon's fields $\clD{1} {}+{} \clD{2}$. The appropriate first
approximation is connected with the radiation of the dyons.
The oscillating bidyon in initial approximation may give rise
to some standing wave in first approximation. According to
properties of the radiation for accelerated charge, an amplitude of this
standing wave is zero at infinity of three-dimensional space.
In this sense we can call it as restricted  standing wave.
The standing wave may change the trajectory $a(x^0)$.
Thus we shall have an energy interchange between the initial approximation
and the standing wave. Some equilibrium oscillating field configuration
may exist for this case.
Because this problem has two-point boundary conditions,
it is possible that the bidyon oscillating system will have some
discrete set of allowable frequencies.

Furthermore, the oscillating bidyon will interact with any plane waves,
which are exact solutions of the model \cite{I1998JHEP}.
The resonance interaction will intercept the plane waves with
bidyon's frequencies. As result we shall have some non-restricted
(in three-dimensional space) standing wave
connected with the oscillating bidyon. Thus it is valid to say that
such oscillating field configuration occupies whole
three-dimensional space.
Of course, near the singular points this configuration has a localization
region, but the associated non-restricted standing wave has some finite
amplitude at infinity.

We can represent the oscillating bidyon solution
by Fourier series in time coordinate.
Thus we have the following series  for the solution
in own coordinate system $y^\mu$:
\begin{eqnarray}
\lb{Sol:BidyonFourierS}
\underline{{\cal D}}
&=& \dsum{\infty}{f {}={} -\infty}
\underline{{\cal D}}_f
\exp\left(-\i f \underline{\omega}\, y^0\right)
\quad,
\end{eqnarray}
where $\underline{\omega} {}\equiv{} \dfrac{2\pi}{T}$,
$\underline{{\cal D}}_f {}={} \underline{{\cal D}}_f (y^i,\underline{\omega})$,
are complex column-function given by the formula
\begin{eqnarray}
\underline{{\cal D}}_f
&=&
\dfrac{1}{2\pi}\,\dint{T}{0}
\underline{{\cal D}}\,
\exp\left(\i f \underline{\omega}\, y^0\right) \,\d y^0
\quad.
\end{eqnarray}
With help Lorentz transformation we can obtain from (\ref{Sol:BidyonFourierS})
 the following moving bidyon solution:
\begin{eqnarray}
\lb{Sol:BidyonFourier}
{\cal D}
&=& \dsum{\infty}{f {}={} -\infty}
{\cal D}_f\,
\exp\left(\i f \Theta\right)
\quad,
\end{eqnarray}
where $\Theta {}={} k_\mu\,x^\mu$, $k_\mu\,k^\mu {}={} -\underline{\omega}^2$,
and the fields $\D_f,\,\B_f$ (for the column ${\cal D}_f$) are  connected
with the fields $\underline{\D}_f,\,\underline{\B}_f$
(for the column $\underline{{\cal D}}_f$)
by formulas of Lorentz
transformation (\ref{Sol:Dyonm}\aref{a}) with functions $\E (\D,\B)$,
$\H (\D,\B)$ (\ref{Expr:EHDB}).

\section{Effect of gravitational interaction}

In the article \cite{I1998JHEP} we had considered
the distortion of a
small amplitude quick-oscillating wave against the background
of a high field, for this nonlinear electrodynamic model.
We had shown that this distortion coincide with light beams distortion
in gravitational field with the following metric:
\begin{equation}
\lb{Def:barg}
\bar{g}^{\mu\nu} {}={} g^{\mu\nu} {}-{} \alpha^2\,T^{\mu\nu}
\quad,
\end{equation}
where $g^{\mu\nu}$ is the metric of basic Minkowskian space and
the energy-momentum tensor $T^{\mu\nu}$ is formulated from
the background field.

Bidyon solution (\ref{Sol:BidyonFourier}) has the quick-oscillating wave part.
 As distinct from
light wave, this wave has the rest frequency $\underline{\omega}$.
But we can presume that this wave (in the region with  small amplitude
of it) will have the analogous behavior that light wave.

Thus let us
consider the problem for propagation of the small amplitude
quick-oscillating wave with the rest frequency
against a high field background. Here we shall use a method that
some distinct from one, which used in the article \cite{I1998JHEP}.
Let us take the cartesian coordinate system $\{x^\mu\}$ with
the metric $g^{\mu\nu} {}={} h^{\mu\nu}$ such that
\begin{equation}
 |h_{00}| {}={}
|h_{11}| {}={} |h_{22}| {}={} |h_{33}| {}={} 1\quad,\qquad
h_{\mu\nu} {}={} 0 \quad \mbox{for}\quad \mu {}\neq{} \nu
\quad.
\end{equation}
In this section we don't define one of two signatures of the Minkowskian
metric: $(-+++)$ or $(+---)$.

In this coordinate system we have the following form of
equation (\ref{Eq:EulerBIA}) for the potential
$A_\mu$ outside of the singular set:
\begin{eqnarray}
\lb{Eq:EulerBIAc}
&&\qquad C^{\mu\nu\sigma\rho}\,
\frac{\p^2 A_\rho}{\p x^\mu \p x^\sigma} {\,}={\,} 0
\quad.
\end{eqnarray}

Let we have a small field configuration  $\underline{\tilde{A}}_\mu (y^\nu)$,
which is periodical on $y^0$ with circular frequency $\underline{\omega}$
and satisfies the Maxwell equations
\begin{eqnarray}
\lb{Eq:EulerBIAMax}
&&\qquad \CMax^{\mu\nu\sigma\rho}\,
\frac{\p^2 \underline{\tilde{A}}_\rho}{\p y^\mu \p y^\sigma} {\,}={\,} 0
\quad,
\end{eqnarray}
where
\begin{eqnarray}
\lb{Def:CMax}
&&
\CMax^{\mu\nu\sigma\rho} {\,}={\,}
h^{\mu\sigma}\,h^{\nu\rho} {}-{}
h^{\mu\rho}\,h^{\nu\sigma}
\quad.
\non
\end{eqnarray}
This field configuration may represent the quick-oscillating part
of the bidyon solution (\ref{Sol:BidyonFourierS}) in the region
where the electromagnetic field is small and the nonlinearity can be neglected.
With help the Lorentz transformation we can obtain also an
appropriate moving field configuration.
But if we have its moving against the
background of a high field, we can presume a change
for parameters of this small field configuration.
Let us take a solution of the model as the background field
$\stackrel{\circ}{A}_\mu$
and a moving field configuration  $\tilde{A}_\mu (x^\nu)$, which is
related with  $\underline{\tilde{A}}_\mu (y^\nu)$ by some
unknown transformation. We substitute their sum
\begin{eqnarray}
A_\mu &=& \stackrel{\circ}{A}_\mu {}+{} \tilde{A}_\mu
\end{eqnarray}
into system of equation (\ref{Eq:EulerBIAc}).
Taking into account that for the associated electromagnetic field
\quad$\left|\vphantom{\stackrel{\circ}{F}_{\mu\nu}}
\tilde{F}_{\mu\nu}\right| {}\ll{}
\left|\stackrel{\circ}{F}_{\mu\nu}\right|$\quad
and
\quad$\left|\vphantom{\dfrac{\p\stackrel{\circ}{F}_{\mu\nu}}{\p x^\delta}}
\dfrac{\p\tilde{F}_{\mu\nu}}{\p x^\delta}\right|\,
\left|\stackrel{\circ}{F}_{\sigma\rho}\right|
{} \gg {}
\left|\dfrac{\p\stackrel{\circ}{F}_{\mu\nu}}{\p x^\delta}\right|\,
\left|\vphantom{\stackrel{\circ}{F}_{\mu\nu}}
\tilde{F}_{\sigma\rho}\right|$\quad\\
we obtain the following equation:
\begin{eqnarray}
\lb{Eq:EulerBIAMaxA}
C^{\mu\nu\sigma\rho}\,
\frac{\p^2 \tilde{A}_\rho}{\p x^\mu \p x^\sigma} &=& 0
\quad,
\end{eqnarray}
where  coefficients $C^{\mu\nu\sigma\rho}$ (\ref{Def:C}) include the
background field $\stackrel{\circ}{F}_{\mu\nu}$ only.

Into some restricted region, where we can believe that the background
field is constant, we can try to find a coordinate transformation
\begin{equation}
\d y^\mu {\,}={\,} G^\mu_{.\nu}\,\d x^\nu
\quad, \qquad
\d x^\mu {\,}={\,} \check{G}^\mu_{.\nu}\,\d y^\nu
\quad,
\end{equation}
which transform system of equations (\ref{Eq:EulerBIAMaxA})
to system (\ref{Eq:EulerBIAMax}). We shall have this transformation if
\begin{equation}
C^{\mu\nu\sigma\rho}  {\,}={\,}
\check{G}^\mu_{.\delta}\,\check{G}^\nu_{.\gamma}\,
\check{G}^\sigma_{.\zeta}\,\check{G}^\rho_{.\xi}\;
\CMax^{\delta\gamma\zeta\xi} {\,}={\,}
\bar{g}^{\prime\mu\rho}\,\bar{g}^{\prime\nu\sigma} {}-{}
\bar{g}^{\prime\mu\sigma}\,\bar{g}^{\prime\nu\rho}
\quad,
\end{equation}
where
$\bar{g}^{\prime\mu\rho} {}\equiv{} \check{G}^\mu_{.\zeta}\,
\check{G}^\rho_{.\xi}\,h^{\zeta\xi}$
 is a metric associated with this transformation.

With help the direct substitution
we obtain that
$\bar{g}^{\prime\mu\rho} {}={} \bar{g}^{\mu\rho}$, where
the metric $\bar{g}^{\mu\rho}$ is defined by formula (\ref{Def:barg})!

Let us write the basic relations, which connect the
metric $\bar{g}^{\mu\rho}$ with the transformation matrixes.
\begin{eqnarray}
\lb{gGGh}
\bar{g}^{\mu\nu}\,G^\sigma_{.\mu}\,G^\rho_{.\nu} &=& h^{\sigma\rho}
\quad,\qquad
\bar{g}^{\mu\rho} {}\equiv{} \check{G}^\mu_{.\zeta}\,
\check{G}^\rho_{.\xi}\,h^{\zeta\xi}
\quad.
\end{eqnarray}

Thus we have a solution
of system (\ref{Eq:EulerBIAMaxA}) into the restricted region
(where ${\bar{g}^{\mu\nu} {}\approx{} \const}$) in the following form:
\begin{eqnarray}
\lb{Sol:tA}
\tilde{A}_\mu &=&
G^\nu_{.\mu}\,\underline{\tilde{A}}_\nu (y^\sigma)
\quad.
\end{eqnarray}

The metric $\bar{g}^{\mu\nu}$, calculated by formula (\ref{Def:barg})
with the background field for full four-dimensional space, define
some curvilinear Riemann space, in general case.
Thus, according to known theorem of Riemann geometry,  we can't
consider $G^{\mu\nu}$ as coordinate transformation tensor for full
space. But we can always make the transformation to the cartesian
coordinates $\{y^\mu\}$ for one point of four-dimensional space.
As this point we take the point $\{\bar{x}^\mu\}$, where
$\bar{x}^i$ are three-dimensional coordinates for some central point
in a localization region of the field $\tilde{A}_\mu$
 at the time moment $\bar{x}^0$.
Thus for using solution (\ref{Sol:tA}) in full four-dimensional
space we must take $G^\nu_{.\mu} {}={} G^\nu_{.\mu}(\bar{x}^\rho)$
 in formula (\ref{Sol:tA}). But because
$\bar{x}^i {}={} \bar{x}^i(\bar{x}^0)$ we have
$G^\nu_{.\mu} {}={} G^\nu_{.\mu}(\bar{x}^0)$ into formula (\ref{Sol:tA}).
For the coordinates $\{y^\mu\}$ we take the following
expressions:
\begin{eqnarray}
\non
y^i &=& G^i_j\left(x^j {}-{} \bar{x}^j \right)
\quad,\qquad G^i_j {}={} G^i_j (\bar{x}^0)
\quad,\qquad \bar{x}^i {}={} \bar{x}^i(\bar{x}^0)
\quad,\\
y^0 &=& \int \bar{k}_\mu\,\d x^\mu
\quad,\qquad \bar{k}_\mu (x^\nu) {\,}\equiv{\,} G^0_\mu (x^\nu)
\quad.
\lb{Def:bark}
\end{eqnarray}
Because the solution $\underline{\tilde{A}}_\nu (y^\sigma)$
is periodical by $y^0$, the proper time $y^0$ play a role of
normalized phase for quick-oscillating wave (\ref{Sol:tA}).
In this case the ordinary phase is $\underline{\omega}\,y^0$ and
the components of wave vector are $\underline{\omega}\,\bar{k}_\mu$.
We have also the following evident relations:
\begin{eqnarray}
\lb{Rel:dkdx}
\frac{\p \bar{k}_\mu}{\p x^\nu} &=& \frac{\p \bar{k}_\nu}{\p x^\mu}
\quad.
\end{eqnarray}

According to formulas (\ref{gGGh}) and definition for $\bar{k}_\mu$
(\ref{Def:bark}) we have
\begin{eqnarray}
\lb{Rel:Disper}
\bar{g}^{\mu\nu}\,\bar{k}_\mu\,\bar{k}_\nu &=& h^{00}
\quad,
\end{eqnarray}
where $\bar{g}^{\mu\nu} {}={} \bar{g}^{\mu\nu}(x^\rho)$,
$\bar{k}_\mu {}={} \bar{k}_\mu (x^\rho)$.\\
This is dispersion relation for the quick-oscillating wave
(see also \cite{I1998JHEP}).

Let us adopt the proper time $y^0 (x^\mu)$ at the point $\bar{x}^\mu$ as
parameter of motion $s$ for the
localization region of the field $\tilde{A}_\mu$.
That is $s {}\equiv{} y^0(\bar{x}^\mu)$.
Thus we have the following definition for four-velocity:
\begin{eqnarray}
\lb{Def:U}
U^\mu &\equiv& \dfrac{\d \bar{x}^\mu}{\d s}
{\,}={\,} \check{G}^\mu_{.0}(\bar{x}^\nu)
\quad.
\end{eqnarray}

Definitions for $U^{\mu}$ (\ref{Def:U}), $\bar{k}_\nu$
(\ref{Def:bark}) and formulas (\ref{gGGh}) give also the following
relations for the point $\{\bar{x}^\mu\}$:
\begin{equation}
\lb{Rel:kU}
\bar{k}_\mu\,U^\mu {}={} 1
\quad,\qquad
U^\mu {}={} h^{00}\,\bar{g}^{\mu\nu}\,\bar{k}_\nu
\quad.
\end{equation}

Let us introduce the inverse tensor $\check{\bar{g}}_{\mu\nu}$
for the tensor $\bar{g}^{\mu\nu}$:
\begin{eqnarray}
\lb{Def:invg}
\check{\bar{g}}_{\mu\nu}\,\bar{g}^{\nu\rho} &=& \delta^\rho_\mu
\quad.
\end{eqnarray}
Of course, within the limits of the introduced  Riemann space the
quantities $\check{\bar{g}}_{\mu\nu}$ are components of the
metric tensor with inferior indexes. But for the basic Minkowskian
space the quantities $\bar{g}_{\mu\nu}$ are components of some
tensor field and $\check{\bar{g}}_{\mu\nu}$ are components of
appropriate inverse one.

From (\ref{Rel:kU}) and (\ref{Def:invg}) we immediately obtain
for the point $\{\bar{x}^\mu\}$ that
\begin{eqnarray}
\check{\bar{g}}_{\mu\nu}\,U^\mu\,U^\nu &=& h^{00}
\quad.
\end{eqnarray}
Thus we have
\begin{eqnarray}
\d s^2 &=& \dfrac{1}{h^{00}}\,
\check{\bar{g}}_{\mu\nu}\,\d\bar{x}^\mu\,\d\bar{x}^\nu
{\,}={\,}
\left|\check{\bar{g}}_{\mu\nu}\,\d\bar{x}^\mu\,\d\bar{x}^\nu\right|
\quad.
\end{eqnarray}

From (\ref{Rel:kU}) we have also
\begin{eqnarray}
\lb{Expr:kgU}
\bar{k}_\mu &=& \dfrac{1}{h^{00}}\,\check{\bar{g}}_{\mu\nu}\,U^\nu
\quad.
\end{eqnarray}

Let us obtain an equation for the trajectory $\bar{x}^\mu (s)$.
Differentiating dispersion relation (\ref{Rel:Disper})
by some coordinate $x^\rho$ and using  relations (\ref{Rel:dkdx})
and expression for $U^\nu$ through $\bar{k}_\mu$ (\ref{Rel:kU})
we obtain for the point $\{\bar{x}^\mu\}$
\begin{eqnarray}
\lb{Eq:Geodtr}
\dfrac{\p \bar{g}^{\mu\nu}}{\p x^\rho}\,\bar{k}_\mu\,\bar{k}_\nu
{}+{} 2\,h^{00}\,U^\nu\,\dfrac{\p \bar{k}_\rho}{\p x^\nu}
{}={} 0
\quad&\Longrightarrow&\quad
\dfrac{\p \bar{g}^{\mu\nu}}{\p x^\rho}\,\bar{k}_\mu\,\bar{k}_\nu
{}+{} 2\,h^{00}\,\dfrac{\d \bar{k}_\rho}{\d s}
{}={} 0
\;\;.
\end{eqnarray}

Substituting (\ref{Expr:kgU}) into (\ref{Eq:Geodtr}) and using
(\ref{Def:invg}) we obtain the trajectory equation in the following
form:
\begin{eqnarray}
\dfrac{\d U^\mu}{\d s} {}+{} \bar{\Gamma}^\mu_{\nu\rho}\,U^\nu\,U^\rho
&=& 0
\quad,
\end{eqnarray}
where
\begin{eqnarray}
\bar{\Gamma}^\mu_{\nu\rho} &=&
\dfrac{1}{2}\,\bar{g}^{\mu\delta}\left(
\dfrac{\p\check{\bar{g}}_{\delta\nu}}{\p x^\rho} {}+{}
\dfrac{\p\check{\bar{g}}_{\delta\rho}}{\p x^\nu} {}-{}
\dfrac{\p\check{\bar{g}}_{\nu\rho}}{\p x^\delta}
\right)
\quad.
\end{eqnarray}
As we see, within the limits of the introduced Riemann space
(for which $g^{\mu\nu} {}={} \bar{g}^{\mu\nu}$,
$g_{\mu\nu} {}={} \check{\bar{g}}_{\mu\nu}$,
$\Gamma^\mu_{\nu\rho} {}={} \bar{\Gamma}^\mu_{\nu\rho} $)
this equation is geodesic line equation!

Notice that here we use the primordial equation for the potential
$A_\mu$ (\ref{Eq:EulerBIAc}) without any gauge condition.
It is evident, we can take the gauge condition
$\bar{g}^{\mu\nu}\,\dfrac{\p\,A_\nu}{\p x^\mu} {}={} 0$ for equation
(\ref{Eq:EulerBIAMaxA}),
which is transformed to the condition
$h^{\mu\nu}\,\dfrac{\p\,\underline{A}_\nu}{\p y^\mu} {}={} 0$
for equation (\ref{Eq:EulerBIAMax}).

Thus we have the Riemann space which is induced by background field.
The relatively small quick-oscillating wave field configuration with rest
frequency and localized near point $\bar{x}^i (\bar{x}^0)$
moves on geodesic lines of this Riemann space.
This looks as motion of gravitating particles.
Thought the bidyon's field is high near the singular points,
we may presume that in some cases the quick-oscillating bidyon
field configuration will move also on these geodesic lines. We shall
discuss this question in the following section.

\section{Nonlinear electrodynamic world with oscillating bidyons}

In this section we shall discuss a possible world constructed from the
great number of the bidyon-type field configurations. It is evident
that in this world the bidyons and its possible
strong coupling combinations may play the role of particles.
Particles with full angular momentum multiples to $\dfrac{\hbar}{2}$
may be constructed from some number of the bidyons.
It is amply clear that moving bidyon  (\ref{Sol:BidyonFourier})
has both particle and wave properties.

Because
$\dfrac{\,\bd\,}{\,\bar{b}\,} {}={} \bar{\alpha} {}\ll{} 1$,
 we may build a perturbation
theory with the fine structure constant $\bar{\alpha}$ as small parameter,
for some aspects of the mathematical
model of this world. Thus we have an analogy with perturbative
procedure of quantum electrodynamics.

The world's field configuration
is solution of the model's equations
in whole four-dimensional space and
each dyon singularity has a determined trajectory providing the
satisfaction to the point boundary conditions. To obtain the solution
for whole space, actually we may use the method of successive
approximation only. According our method describing in
section \ref{Sec:PertMethod}, the dyon's trajectories are corrected
for each step of iteration with help of some integral conditions.
In the present article we use the integral conservation law of
the energy-momentum for this purpose. Thus this method for obtaining
the trajectories is non-local in character. We have the classical
generalized Lorentz force in initial approximation,
 for the case when we may consider
the conservation law for the localization region of one dyon only.
In general case the integral conditions may give the following effect.
If we have a change of the field near some distant (from
the trajectory) point, then we may obtain a change of this trajectory
(without a propagation of disturbance from this distant point).
This looks as quantum behavior when events at space separated points
behave as though there is an instantaneous interaction between the
events. According our concept, this pattern may be
connected with the specific iterative way on which we go
(in our calculations)
to unknown world's field configuration in space-time.

The wave vector $k_\mu$ of
 moving quick-oscillating bidyon (\ref{Sol:BidyonFourier})
(without any external given fields)
satisfies the following dispersion relation:
\begin{eqnarray}
\lb{Rel:dispk}
k^\mu\,k_\mu &=& h^{00}\,\underline{\omega}^2
\quad.
\end{eqnarray}
We have also
the following relation for the vector of full
energy-momentum of the bidyon field configuration:
\begin{eqnarray}
\lb{Rel:dispp}
P^\mu\,P_\mu &=& h^{00}\,m^2
\quad,
\end{eqnarray}
where $m$ is the full rest energy interpreted as mass.\\
However, because we have the non-restricted wave
(with infinite three-dimensional energy integral) as a part of
the bidyon configuration, how we must calculate the full energy-momentum
in this case is not clear. A future consideration of the problem for motion
of the bidyon with the associated quick-oscillating waves in given external
field must give a value of the mass.
We can denote the relation between the mass
and the rest frequency  as
${\cal C} {}\equiv{}
m {}/{} \underline{\omega}$.
Thus we have
\begin{eqnarray}
P_\mu &=& {\cal C}\,k_\mu
\quad.
\end{eqnarray}

To correlate between the bidyons (or multidyons) and real
particles, we must have that ${\cal C} {}={} \hbar$.
To determine truth or falsity of this
condition for the bidyon solution, we must have this
solution  and value of the mass.
But the behavior of bidyon's initial approximation (\ref{Sol:InitBidyon})
at very small interdyonic distances (high energies)
may suggest some optimism (see (\ref{TrajectIdeal}),(\ref{Expr:omegaappr1})).
Of course, outside of the regions immediately adjacent
to the singular points the associated exact solution may be
significantly different from the initial approximation. Thus this
problem need further consideration.

Gravitational interaction (in the sense of bimetric gravitational
theory) may exist for the bidyon solution.
The quick-oscillating part of the solution with small amplitude
will have tendency to move on geodesic line of induced by
background field Riemann space. We may presume that in some cases
this tendency will be a controlling factor for motion of the bidyon,
that is the bidyon will be piloted by its small amplitude
quick-oscillating part.
This case may takes place when we have the quick-oscillating
background field such that the classical electromagnetic interaction
for the bidyon is absent because of averaging. We must presume also that
the averaging cancels the effects of associated resonance interaction.
But induced metric (\ref{Def:barg}) includes squares of the
field and the appropriate effect will not wipe out by averaging.

Thus we may hope that the nonlinear electrodynamic world will
have a correlation with the real material world.
In this case we could call an associated theory as
{\it Nonlinear Electrodynamic Theory of the World}.
Such theory may be considered as like Einstein's unified field theory
with electromagnetic field as basic. In the context of this approach,
all diversity of solutions and nonlinear effects
for the electrodynamic field model must correlate with
all real particles and their interactions. The evident practical
significance of such approach consists in a possible
discovery any new real effects with help of the theoretical investigation
of the model.
But whilst still we have more questions than answers on this
arduous but feasible way.

\section{Conclusions}

Thus we have presented the initial theory for nonlinear
Born-Infeld electrodynamics with dyonic singularities.
We have proposed the method for investigation of interaction
between the dyons. For a special case we have obtained the generalized
Lorentz force acting on the dyon. We have introduced the concept
of bidyon, interpreted as electromagnetic model of particle with spin,
and we have investigated its initial approximation.
We have obtained the effect of gravitational interaction, which appear
within the limits of this electrodynamic model.
We have also discussed possible correlations between
the nonlinear electrodynamic world and real material world.

\bref
\bibitem{I1992} A.A.~Chernitskii,
Long-range interaction of four-vector field solitons of
the Min\-kows\-kian space,
{\it Theor. Math. Phys.} {\bf 90}(3) (1992) 260 (380 in Russian).
\bibitem{I1995GR14} A.A.~Chernitskii,
Gravitation as long-range
interaction of solitons in non-linear electrodynamics,
{\it Book of abstracts GR14 (Florence)} (1995), {\bf A}96. .
\bibitem{I1998JHEP} A.A.~Chernitskii,
Light beams distortion in nonlinear electrodynamics,\\
{\it J. High Energy Phys.} 11 (1998) 015 [hep-th/9809175].
\bibitem{BornInfeld} M.~Born and L.~Infeld, Foundations of the New
Field Theory,\\ {\it Proc. Roy. Soc.} {\bf 144} (1934) 425.
\bibitem{Eddington1924} A.S.~Eddington, {\it The Mathematical Theory
of Relativity} (Cambridge, 1924).
\bibitem{I1998HPA} A.A.~Chernitskii,
Nonlinear electrodynamics with singularities
(Modernized Born-Infeld electrodynamics), {\it Helv. Phys. Acta}
{\bf 71} (1998) 274--287 [hep-th/9705075].
\bibitem{Schwinger} J.~Schwinger, A Magnetic Model of Matter,
 {\it Science} {\bf 165} (1969) 757--761.
\bibitem{BialynickiBirula} I. Bialynicki-Birula,
Nonlinear Electrodynamics: Variations on a Theme
fy Born and Infeld, {\it Quantum Theory of Particles and Fields}
ed.B.~Jancewicz and J.~Lukierski, (1983) 31-47.
\bibitem{Gibbonshepth9506035} G. Gibbons,
Electric-Magnetic Duality Rotations in Non-Linear Electrodynamics,
{\it Nucl. Phys.} {\bf B}454 (1995) 185-206
[hep-th/9506035].
\bibitem{Dirac1948} P.A.M. Dirac, The theory of magnetic poles,
{\it Phys. Rev., Sec. ser.} {\bf 74} (1948) 817-830.
\bibitem{CurantHilbert} R. Curant and D. Hilbert, {\it Method der Mathematical
Physik},\\ (Verlag von Julius Springer, Berlin, 1931).
\bibitem{HustonPym} V.C.L.~Huston and J.S.~Pym,
{\it Applications of Functional Analysis and Operator Theory}
(Academic Press, London, 1980).
\eref

\end{document}